\documentclass[12pt,letterpaper,aps,nofootinbib,floats,floatfix]{revtex4}
\usepackage{graphicx,xspace}
\usepackage{hyperref}
\begin{document}

\newcommand {\be} {\begin{equation}}
\newcommand {\ee} {\end{equation}}
\newcommand {\bea} {\begin{eqnarray}}
\newcommand {\eea} {\end{eqnarray}}
\newcommand {\eq} [1] {Eq.\ (\ref{#1})}
\newcommand {\eqs} [2] {Eqs.\ (\ref{#1}) and (\ref{#2})}
\newcommand {\Eq} [1] {Equation (\ref{#1})}
\newcommand {\Eqs} [2] {Equations (\ref{#1}) and (\ref{#2})}
\newcommand {\fig} [1] {Fig.\ \ref{#1}}
\newcommand {\figs} [2] {Figs.\ \ref{#1} and \ref{#2}}
\newcommand {\Fig} [1] {Figure \ref{#1}}
\newcommand {\Figs} [2] {Figures \ref{#1} and \ref{#2}}

\def \eps {\epsilon}
\def \veps {\varepsilon}
\def \pl {\partial}
\def \hf {{1 \over 2}}
\def \mf {\mathbf}
\def\figdir{}
\def\sss{\scriptscriptstyle}

\def\lsim{\mbox{\raisebox{-.6ex}{~$\stackrel{<}{\sim}$~}}}
\def\gsim{\mbox{\raisebox{-.6ex}{~$\stackrel{>}{\sim}$~}}}


\title{\bf How Delicate is Brane-Antibrane Inflation?}

\author{
\normalsize Loison~Hoi and James~M.~Cline}
\affiliation{
Department of Physics, McGill University\\
3600 University Street, Montr\'eal, Qu\'ebec, Canada H3A 2T8\\
E-mail: hoiloison@physics.mcgill.ca, jcline@physics.mcgill.ca}
\date{11 May 2009}

\begin{abstract}

We systematically explore the parameter space of the state-of-the-art
brane-antibrane inflation model (Baumann {\it et al}., 
arXiv:0706.0360, arXiv:0705.3837)  which is one of the most
rigorously derived from string theory, applying the cosmic background explorer
normalization and constraint on the spectral index.  We improve on
previous treatments of uplifting by antibranes and show that the
contributions from noninflationary throats play an important role in
achieving a flat inflationary potential.  To quantify the degree of
fine-tuning needed by the model, we define an effective volume in the
part of parameter space which is consistent with experimental
constraints, and using Monte Carlo methods to search for a set of
optimal parameters,  we show that the degree of fine-tuning is
alleviated by 8 orders of magnitude relative to a fiducial point
which has previously been considered. 
In fact, close to the optimal
parameter values, fine-tuning is no longer needed for any of the
parameters.  We show that in this natural region of the parameter
space, larger values of $n_s$ close to $0.99$ (still within 2$\sigma$
of the WMAP5 central value) are favored, giving a new aspect of
testability to the model.
\end{abstract} 

\maketitle

\section{Introduction}

Brane-antibrane inflation is one of the most distinctive and highly 
developed applications of string theory to cosmology.  Although the
basic idea seems simple---getting inflation from the potential between a
brane and antibrane which are separated along an extra
dimension---making it work has proved to be quite challenging.  The
earliest versions were incomplete due to the lack of an explicit
mechanism for stabilizing the moduli, notably those associated with
the size and shape of the extra dimensions.  Major progress in this
respect was made in Ref.\ \cite{KKLMMT}, where  warped
compactification of type IIB string theory using the
Klebanov-Strassler throat \cite{KS} was exploited.

In Ref.\ \cite{KKLMMT}, it was shown that the potential for a brane
falling down the throat suffers from the $\eta$ problem: the
curvature of the inflaton potential is generically of order $H^2$,
making it too steep for inflation.  To counteract this, the
generically large inflaton mass must be tuned to a small value using
a canceling contribution from the $F$-term potential, coming from the
superpotential 
\be 
	W =  W_0 + A_0(\phi) e^{-a T},
\ee 
where $\phi$
denotes the brane position (inflaton) and $T$ is the complex K\"ahler modulus,
whose real part $\sigma$ determines the overall volume of
the extra dimensions.  The second term  arises
from nonperturbative physics like gaugino condensation on a D7-brane,
which is one of the essential ingredients of the compactification.
The existence of the $F$-term potential was already known to be
necessary for stabilizing $T$ \cite{KKLT}; however the dependence of
$A_0$ on $\phi$ was not known by the authors of Ref.\ \cite{KKLMMT}; they
merely parametrized this dependence.

The actual dependence $A_0(\phi)$ however is calculable within string
theory, and this missing step was carried out in Ref.\ 
\cite{Baumann:2006th}.  It was subsequently shown \cite{BCDF} that the
explicit form of $A_0(\phi)$ was apparently not amenable to achieving the
desired cancellation to get a small inflaton mass. However this
conclusion depended on exactly how the D7-brane was embedded in the
throat.  A different embedding was considered in Refs.\
\cite{Krause,Baumann:2007np, Baumann} which yielded the desired form of
the $F$-term potential.  With sufficient tuning of parameters, it was
possible to achieve inflation.  Because of this need for substantial
tuning, the scenario was dubbed ``delicate" in Refs.\
\cite{Baumann:2007np,Baumann}. The difficulties were further elaborated 
in Ref.\ \cite{Panda}, which surprisingly concluded that it was difficult
to satisfy both the cosmic background explorer (COBE) normalization\footnote{The
historical term ``COBE normalization" refers to the value of the
primordial power spectrum at $k=0.002\ {\rm Mpc}^{-1}$; of course, 
we use the latest Wilkinson Microwave Anisotropy Probe (WMAP) value.} and the constraint on the spectral index. 
Normally one expects that there is enough freedom to adjust the overall 
scale of the potential to achieve the normalization of the amplitude,
and indeed, we will show that satisfying the normalization presents no
extraordinary new difficulty.\footnote{We will see in Section
\ref{varyingA0} that the normalization and the spectral index are
correlated; hence further fine-tuning is needed to construct a model
satisfying the current observational constraints.}

On the other hand, we do point out an additional challenge which was
not consistently treated in previous work.   Namely, it is necessary for the
cosmological constant to vanish at the end of inflation.  In Ref.\
\cite{Baumann}, sufficient flatness of the potential was achieved by
tuning the parameter $s$, which is the ratio of the uplifting energy
to the absolute value of the negative energy of the anti-de Sitter (AdS) minimum in
the absence of uplifting.  In fact, this ratio is fixed (to
approximately unity) by the necessity to uplift to a Minkowski
minimum, thereby removing it as a tunable parameter. We take
advantage of the fact that its tunability can be restored by
considering the contributions from other throats to the uplifting.   

In Refs.\ \cite{Baumann:2007np,Baumann}, it is claimed that the
brane-antibrane inflationary model must be rather finely tuned.  However,
there is little quantitative basis for this statement.  Part of our
objective is to invent a quantitative measure of the degree of tuning. 
We propose a measure of tuning via the relative volume of parameter space around a
given point, which is consistent with inflation.  This statistic allows
us to say to what fractional precision the parameters have to be adjusted
in order to get inflation consistent with observations; e.g.,
parameter $x$ must be tuned to one part in 100, parameter $y$ to one part in 1000,
{\it etc}.  Using this measure, we show that indeed, the model is quite
fine-tuned at the parameter values which have been previously suggested.
However, we will show that there exists a more favorable set of
parameters around which the tuning problem is much less severe, by
adapting Monte Carlo methods which have been widely used in other
cosmological applications for this purpose. 
We also consider the degree of fine-tuning using an
alternative measure---the absolute volume of the allowed parameter
space; we compare the two measures and show that the improvements in
fine-tuning are of the same order.

A further delicate aspect of the model is the problem of
overshooting the flat part of the potential by starting too high
\cite{Underwood:2008dh}.   With the wrong initial conditions, the
inflaton gathers too much speed to roll slowly in the flat region of
the potential. We investigate the scope of initial conditions which
is compatible with inflation and show that this aspect of the tuning
problem is also improved somewhat in the most favorable parameter
range.

A final technical issue concerns the extent to which the model can be
treated in terms of a single field, due to the fact that the
trajectory curves in the space of $\phi$, the brane modulus, and 
$T$, the K\"ahler modulus.   A seemingly reasonable treatment,
involving both fields plus an analytic approximation for the
dependence of $T$ on $\phi$, can result in an incorrect prediction
for the spectral index.   We show that the single-field approximation
can be a good one, provided one is sufficiently careful.

\section{The Potential of $D$-Brane Inflation}

\begin{figure}[htp]
\centering{\includegraphics[width=3in]{\figdir 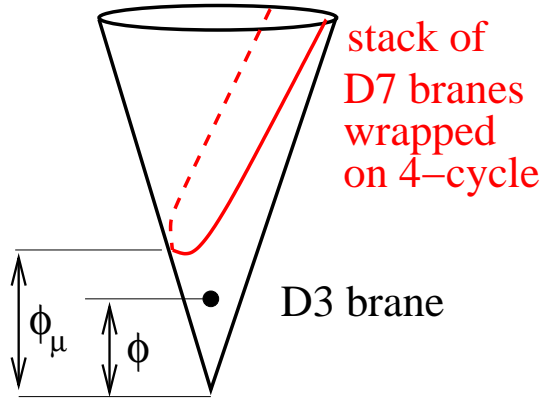}}
\caption{Schematic representation of D7-brane stack and D3-brane in
Klebanov-Strassler throat.}
\label{throat}
\end{figure}

The inflaton potential derived by Baumann {\it et al}.\ \cite{Baumann:2007np,
Baumann} consists of two contributions: a supersymmetric (SUSY)
$F$-term potential $V_F$, and a SUSY-breaking contribution due to
anti-D3-branes, $V_D$, which is necessary for uplifting the minimum
from a negative value (an AdS minimum) to zero.   $V_F$
depends on the correction to the nonperturbative superpotential, 
\be
	A_0(\phi) = A_0\, g^{1/n}(\phi) \equiv A_0 
	\left[1 + \left(\phi\over\phi_\mu\right)^{3/2}\right]^{1/n}.
\ee
$A_0$ is now taken to be a constant, 
$n$ is the number of D7-branes in the stack, and $\phi_\mu$ is
proportional to the radial position where the D7-brane stack
comes closest to the bottom of the throat.  For $\phi <
\phi_\mu$, the D3-brane is below the D7-brane stack in the throat;
this is the region where inflation takes place.  The configuration
is illustrated in \fig{throat}. 

The dynamical degrees of freedom are expressed in terms of 
the dimensionless rescalings of the D3-brane
modulus $\phi$ and the K\"ahler modulus $\sigma$, 
\bea
  x &\equiv& {\phi \over \phi_\mu},\\
  \omega &\equiv& a\sigma \equiv {2\pi\over n}\sigma.
\eea
The 
potential is then\footnote{We set the reduced Planck mass to $M_{\rm
Pl} = 1$.}
\bea
  V(x,\omega) &=& V_F(x,\omega) + V_D(x,\omega),\\
  V_F(x,\omega) &=& \frac{a|A_0|^2}{3U^2(x,\omega)}e^{-2\omega}g^{2/n}\left[2\omega + 6 - 6\left|{W_0 \over A_0}\right|e^\omega g^{-1/n}+\frac{3}{ng}\left({cx \over g}-x^{3/2}\right)\right],\\
  V_D(x,\omega) &=& \frac{D(x)}{U^2(x,\omega)},
\eea
where $c = 9/(4n\omega_0\phi_\mu^2)$, and $\omega_0$ is defined to
be the stable value of $\omega$ when $x=0$, i.e.,
\be
	\left.{\partial V \over \partial\omega}\right|_{x=0,\omega=\omega_0} = 0.
\ee
Thus $\omega_0$ is not a free parameter. 
(See Appendix \ref{omega*} for details.)
The functions $U$ and $D$ come from the K\"ahler potential and the
brane-antibrane Coulombic interaction, respectively, and 
are given by 
\bea
  U(x,\omega) &=& {1 \over a}\left(2\omega - {1 \over 3}\omega_0\phi_\mu^2x^2\right),\\
  D(x) &=& \frac{D_0}{1+C_D\begin{displaystyle}
{D_0 \over x^4}\end{displaystyle}} + D_1,\qquad
C_D = {27 \over 64\pi^2\phi_\mu^4}.
\label{uplift}
\eea
The uplifting term $D(x)$ is generalized relative to
Ref.\ \cite{Baumann} by including an extra contribution $D_1$
representing the contribution to the uplifting which remains
after the brane-antibrane annihilation.\footnote{The revised versions of Ref.\ \cite{Baumann} have $D_{\rm other}$ in Eq.\ (2.20) which plays the role of $D_1$. However, the authors still vary the total uplifting to tune their potential, rather than the ratio of the upliftings. Comparing Eq.\ (3.19) in Ref.\ \cite{Baumann} to \eq{s}, you will see that we allow the ratio of the upliftings to change.} The first
term involving the warped antibrane tension $D_0$ in the inflationary
throat has been resummed rather than Taylor expanded  in the Coulomb
interaction.  The difference between these two forms is negligible in
the slow-roll region of the potential, but the resummed  form has
good behavior as $x\to 0$, unlike the Taylor-expanded version, and
this is convenient for numerical evolution going to the end of
inflation.  (In the region where the potential is flat, the Coulomb
interaction typically plays an unimportant role, and it is sufficient
to approximate $D \simeq D_0 + D_1$.)  The resummed form resolves an
inconsistency in the treatment of Ref.\ \cite{Baumann},
which relies upon the value of the potential at $x=0$; however this
only exists in the expanded form if one ignores the Coulombic part
altogether.

Since the $x\to 0$ limit of the uplifting term is not treated
consistently in most of the existing literature, it is worthwhile to
give some further explanation.\footnote{See the appendix of Ref.\
\cite{CS} for a derivation.}  In a more realistic description, we
would need to include the tachyon field of the brane-antibrane
system, whose mass squared becomes negative when the separation
between the two falls below some critical value of order of the string
length scale. Instead of $x\to 0$, the evolution would continue in
the tachyonic direction.  For our purposes, these details are not
important because inflation has already ended by this time.  What is
important however is to know how much the potential decreases between
inflation and the minimum of the potential, and the latter must be at
$V=0$ so that there is no residual cosmological constant.  Our
resummed expression, which goes to zero as $x\to 0$, correctly models
the fact that the tension of the brane-antibrane system is precisely
the amount by which $V$ changes between inflation and the minimum
\cite{Sen}. Although Ref.\ \cite{Sen} worked in an unwarped
background, it is obvious that even in the warped case, the uplifting
provided by an antibrane must precisely vanish once it is annihilated
by a corresponding brane.  This argument shows that our new parameter
$D_1$ represents the residual tension of the branes left over after
the annihilation, assuming they are in different throats from that of
the inflationary brane.  If they were in the same throat, $D_1$ would
have to be an integer multiple of $D_0$, counting the number of
antibranes in the stack which remains after annihilation of the
mobile brane; this would reduce our ability to tune parameters.  
[Also $C_D$ in \eq{uplift} would have the extra factor
$1+D_1/D_0$ due to the Coulombic attraction between the mobile brane
and the antibranes in the stack.]  On the other hand, antibranes in
separate throats have a tunable tension, via the warp factors of the
throats, since it is the warped tension which appears in the
potential.\footnote{Of course the warp factors themselves are
determined by exponentiated ratios of integers; we assume these can be
approximated by continuously varying warp factors.}

In addition to the D3-brane's radial position in the throat, there
are five angular directions, as well as the imaginary (axionic)
component of $T$.  Some of these have large masses and have been set
to the values which minimize their potential. Others have a nearly
flat potential due to approximate isometries of the throat geometry,
but as usual, such compact directions cannot give rise to any
significant amount of  inflation because their motion is quickly
Hubble damped.  (For interesting possible effects of these fields on
the generation of density perturbations at the end of inflation, see
Ref.\ \cite{shiu}.) On the other hand, the K\"ahler modulus $\omega$
can undergo significant evolution, so it is important to keep both
the $\omega$ and $x$ fields at the outset, even though only one
linear combination is light at any point along the  inflationary
trajectory.


We see that the potential depends on six parameters, $A_0$,
$\phi_\mu$, $n$, $W_0$, $D_0$, and $D_1$.  
To make closer
contact with Refs.\ \cite{Baumann, Panda}, we will use
a different parametrization in place of $W_0$ and $D_1$.   
The stable value of the K\"ahler modulus before uplifting, 
$\omega_F$, is defined through
\be
  \left.\frac{\partial V_F}{\partial \omega}\right|_{x=0,\omega=\omega_F} = 0,
\ee
which gives 
\be
  \label{W0}
  3\left|{W_0 \over A_0}\right|e^{\omega_F} = 2\omega_F + 3.
\ee
Therefore $W_0$ can be traded for $\omega_F$.  In place of the
parameter $D_1$,
we define the ratio of $V_D$ and $V_F$:
\be
  \label{s}
  s \equiv \frac{V_D(0,\omega_F)}{|V_F(0,\omega_F)|},
\ee
which gives
\be
  \label{D1}
  D_1 = \frac{2}{3}a|A_0|^2s\omega_Fe^{-2\omega_F}.
\ee

Now, the potentials are
\bea
  V_F &=& {a|A_0|^2 \over 3U^2} e^{-2\omega}g^{2/n}\left[2\omega+6-2(2\omega_F+3)e^{\omega-\omega_F}g^{-1/n} + {3 \over ng}\left({cx \over g}-x^{3/2}\right)\right],\\
  \label{VD}
  V_D &=& {2a|A_0|^2 \over 3U^2} s\omega_F{\rm
e}^{-2\omega_F}\left(1+\frac{D_{01}}{1+C_D\begin{displaystyle}{D_1
D_{01} \over x^4}\end{displaystyle}}\right),
\eea
where $D_{01}$ is the ratio of two tensions:
\be
  D_{01} \equiv {D_0 \over D_1}.
\ee
As a result, the potential is determined by the following six independent parameters:\footnote{In Section \ref{uplifting} we will show that $s$ is a function of $\omega_F$ when the requirement of uplifting the potential is imposed.}
\be
  s, A_0, D_{01}, \phi_\mu, \omega_F, n.
\ee
The dependent parameters $a$, $c$, and $C_D$ have been defined above.

These parameters of the potential can be expressed in terms of more
fundamental string theoretic quantities \cite{Baumann}, which
imply constraints on their allowed values.   In particular,
$\phi_\mu$ and $\omega_F$ can be expressed through stringy
parameters:
\bea
  \label{phimu}
  \phi_\mu &=& \frac{2}{Q_\mu\sqrt{B_6N_5}},\\
  \omega_F &\simeq& {3N_5 \over 2n} B_4\ln Q_\mu,
\eea
where $N_5>1$ is the 5-form flux (equal to the product of the 3-form
fluxes), $B_6>1$ is the ratio of the bulk part of the 6D volume to
the part due to the throat, $B_4>1$ is the corresponding ratio for
the 4-cycle volume on which the D7-brane is wrapped, and $Q_\mu>1$ is
the ratio of the radial coordinate $r_{\rm UV}$ at the top of the throat
to the value $r_{\mu}$ corresponding to $\phi_\mu$ in \fig{throat}.
(See Appendix \ref{microscopic} for the complete list of theoretical
constraints on the microscopic parameters.) We will take these
constraints into account when we search the parameter space.

In addition to the above, there is another important constraint which
was not discussed by previous papers.  The parameter $A_0$ is related
to the scale $\Lambda$ of gaugino condensation by $A_0 = \Lambda^3$
(see, for example, Ref.\ \cite{KKLT}).  This should certainly be less
than the Planck scale, so one should at least demand that $A_0 < 1$,
and more realistically it should be even smaller.  Below we will find
a  preferred value of $A_0 \simeq 0.01$. 

\begin{figure}[htp]
\centering{\includegraphics[width=4in]{\figdir 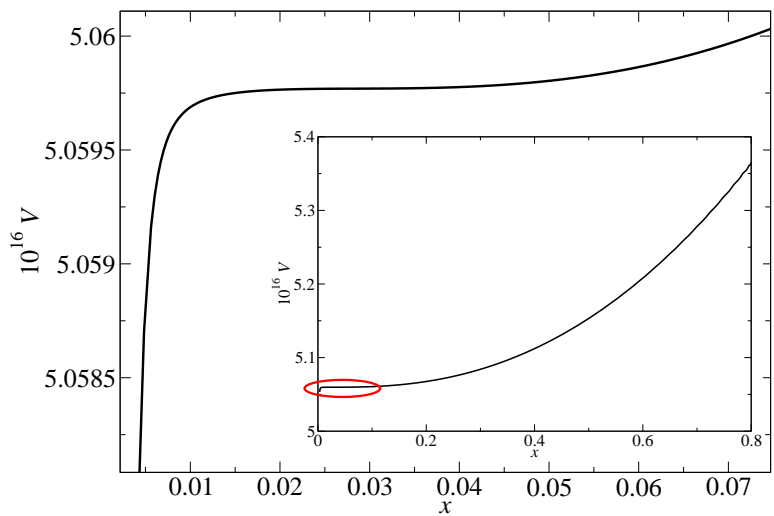}}
\caption{The potential for the optimal parameter set (\ref{optimal})
(to be discussed later), in the vicinity of the inflection point. 
The inset shows the potential over a larger range of $x$.}
\label{potshape}
\end{figure}

The shape of the potential is shown in \fig{potshape}, for
the optimal parameter set (\ref{optimal}) which will be discussed
later.  Away from the inflection point near $x=0.03$, it becomes
steep, and starting from very large values of $x$ can lead to
an overshoot of the flat region.

\section{Equations of Motion}
\label{eom}

The kinetic term of the D3-brane comes from the Dirac-Born-Infeld (DBI) action, of the
form $-a^4T_3\sqrt{1-a^{-4}\dot X^2}$, where $a$ is
the warp factor in the throat.  For small enough field velocities, we
can expand to leading order in $\dot X^2$ to obtain a conventional
kinetic term.  We will check the validity of the approximation
$a^{-4}\dot X^2\ll 1$ in our numerical analysis. In the framework of
the low-energy supergravity effective action, the kinetic term for
the D3-brane and the K\"ahler modulus comes with a nontrivial metric
$G_{I\bar J}$ on the space of the complex fields, taking the form:
\be 
   {\cal L}_{\rm kin} = G_{I\overline J}\partial_\mu
	\Phi^I\partial^\mu {\overline\Phi}^J. 
\ee
This leads to the kinetic term:
\be
  {\cal L}_{\rm kin} = {1 \over a^2U^2}(3{\dot\omega}^2+2\omega_0\omega\phi_\mu^2{\dot x}^2).
\ee
We define the canonical momenta:
\bea
  \pi_x &\equiv& {\partial {\cal L} \over \partial \dot x} = \frac{4\omega_0\omega\phi_\mu^2}{a^2U^2}\dot x,\\
  \pi_\omega &\equiv& {\partial {\cal L} \over \partial \dot \omega} = \frac{6}{a^2U^2}\dot\omega,
\eea
which can be solved for the field velocities,
\bea
  \label{pix}
  \dot x &=& \frac{a^2U^2 }{4\omega_0\omega\phi_\mu^2} \pi_x,\\
  \label{piw}
  \dot \omega &=& {a^2U^2 \over 6} \pi_\omega.
\eea
The equations of motion for the fields are
\bea
  \dot\pi_i + 3H\pi_i &=& {\partial \over \partial\phi_i}\left({\cal L}_{\rm kin}-V\right),\hskip0.5in i = x,\omega,\\
  \label{fieldmann}
  3 H^2 &=& {\cal L}_{\rm kin} + V.
\eea
It is convenient to use the number of $e$-foldings instead of time as 
the independent variable:
\be
  {d \over dt} = H{d \over dN}.
\ee
Then the equations of motion, in suitable form for numerical
integration, are
\bea
  {d\pi_x \over dN} &=& -3\pi_x + {1 \over H}\left(\frac{4\omega_0\phi_\mu^2x}{3aU}{\cal L}_{\rm kin} - {\partial V \over \partial x}\right),\\
  {d\pi_\omega \over dN} &=& -3\pi_\omega + {1 \over H}\left(-{4 \over aU}{\cal L}_{\rm kin} + \frac{2\omega_0\phi_\mu^2{\dot x}^2}{a^2U^2} - {\partial V \over \partial \omega}\right),\\
  {dx \over dN} &=& {1 \over H}\dot x,\\
  {d\omega \over dN} &=& {1 \over H}\dot \omega,
\eea
where $\dot x$ and $\dot \omega$ are determined through
\eqs{pix}{piw}; the derivatives of the potential are given in
Appendix \ref{potapp}.

In the regime that we are interested in, near the tip of the throat, 
$\omega \gg x$ and $\omega\simeq\omega_0$; then  the equations of motion
can be approximately written in canonical form \cite{Panda}:
\bea
  {\cal L}_{\rm kin}  &=& \hf{\dot\phi}^2 + \hf{\dot\chi}^2,\\
  \ddot\phi + 3H\dot\phi &=& - V_\phi,\\
  \ddot\chi + 3H\dot\chi &=& - V_\chi,
\eea
where
\be
  \chi = \sqrt{3 \over 2}\ln\omega,
\ee
and $V_X$ denotes $\partial V/\partial X$. 
Although we will solve the exact equations of motion in this paper,
the canonical fields are useful when discussing
the primordial power spectrum and the spectral index, as will be seen in
the following sections.

\section{The Primordial Power Spectrum}
\label{pr}

The primordial power spectrum for single-field inflation is
\be
  {\cal P_R} = {H^4 \over 4\pi^2{\dot\phi}^2}.
\ee
Generalizing to the noncanonical kinetic term, it can be written as \cite{Cline:2006hu}
\be
  \label{prkin}
  {\cal P_R} = {H^4 \over 8\pi^2{\cal L}_{\rm kin}}.
\ee 
The spectral index is
\bea
  \label{ns}
  n_s \equiv 1 + {d\ln {\cal P_R} \over d\ln k} = 1 + \left(4{d\ln H \over dN} - {d\ln {\cal L}_{\rm kin}\over dN}\right)\left(1 + {d\ln H \over dN}\right)^{-1},
\eea
where $dH/dN$ and $d{\cal L}_{\rm kin}/dN$ can be calculated through
the potential and its derivatives (see Appendix \ref{pwspt} for
details).
\Eq{prkin} is justified when there is effectively just a single field
contributing (single direction in field space), since it is invariant
under field redefinitions. However, it is not valid when the entropy
perturbations make a significant contribution to the power spectrum.

The power spectrum and the spectral index can be expressed in terms of the
slow-roll parameters by defining the adiabatic direction $\psi$ 
in field space, tangent to the inflaton trajectory \cite{Gordon:2000hv}, 
\be
  \dot\psi = \dot\phi\cos\theta + \dot\chi\sin\theta,
\ee
where $\phi$ and $\chi$ are assumed to be canonically normalized, and
\bea
  \cos\theta &=& {\dot\phi \over \sqrt{\dot\phi^2+\dot\chi^2}},\\
  \sin\theta &=& {\dot\chi \over \sqrt{\dot\phi^2+\dot\chi^2}}.
\eea
If $\dot\theta\not=0$, then the trajectory is curved and 
entropy perturbations can source curvature perturbations on 
large scales ($k\to0$).  However if $\dot\theta$ is very small, or
if the entropy mode is suppressed by large curvature of the potential
in the direction orthogonal to $\psi$, which is the case in the
present model \cite{Panda}, the entropy mode can be ignored and then
we have the usual formulas in terms of $\psi$, 
\bea
  \label{prsr}
  {\cal P_R} &=& {1 \over 24\pi^2}{V\over\eps_\psi},\\
  \label{nssr}
  n_s &=& 1 - 6\eps_\psi + 2\eta_{\psi\psi}.
\eea
The definitions of the slow-roll parameters are given in Appendix
\ref{srapp}. 
We will see that in the regime we are interested in, the slow-roll
approximation is well satisfied, hence \eqs{prkin}{prsr} agree with
each other.

\section{Tuning of Parameters}

Using numerical integration and Monte Carlo techniques, we have
undertaken a systematic study of the inflationary dynamics over the
full parameter space of the model. In the following sections, we
will first reproduce the known result \cite{Baumann, Panda} that,
while keeping other parameters fixed,  the tuning of the uplifting
parameter $s$ allows one to obtain inflation with a sufficient number
of $e$-foldings, $N\ge50$.  We then show that  by varying the amplitude
of the nonperturbative superpotential, $A_0$, one can satisfy both
the COBE normalization and the WMAP constraint on the spectral index.
(Recall that  Ref.\ \cite{Panda} claimed that it was difficult to
satisfy both.) We next point out that the $s$ parameter is actually
already fixed by the requirements of uplifting; however, there
remains sufficient freedom to get a flat potential by varying
$D_{01}$ and $\omega_F$.

\subsection{Varying the tension to get flat potential}
\label{varyings}

References \cite{Baumann, Panda} showed that by varying the value of
$D_0$ (proportional to the warped D3-brane tension at the 
tip) one can obtain sufficiently many $e$-foldings of inflation.
Here we further investigate the dependence of the potential on $s$,
which is related to $D_1$ through \eq{D1}, and fix
other parameters as given in Ref.\ \cite{Baumann}:
\be
  A_0 = 1,\ n = 8,\ B_4 = 9.15,\ B_6 = 1.5,\ Q_\mu= 1.2,\ N_5 = 32.
\label{fid1}
\ee
These values imply that 
\be
  \phi_\mu = 0.2406,\ \omega_F = 10.009.
\label{fid2}
\ee
We set our new parameter $D_{01}$ to $1$ for the moment for 
definiteness.\footnote{This is the default parameter set in this and the following two subsections; however, we allow $A_0$ to vary in Section \ref{varyingA0} and $(A_0,D_{01})$ to vary in Section \ref{uplifting}.}
We use initial conditions:
\be
   x_i=0.8,\ \omega_i = \omega_\ast(x_i),\ 
\dot x_i = \dot \omega_i=0,
\ee
where $\omega_\ast$ is the instantaneous minimum which satisfies
\be
  \left.\frac{\partial V}{\partial \omega}
\right|_{\omega_\ast=\omega_\ast(x)}=0.
\label{wstar}
\ee
The initial value $x_i=0.8$ is sufficient for getting inflation while
avoiding the overshoot problem which we will discuss in Section
\ref{overshoot}.  Taking the initial velocities to be zero is
justified since any nonzero values would be quickly Hubble
damped.  

\begin{figure}[htp]
\centering{\includegraphics[width=4in]{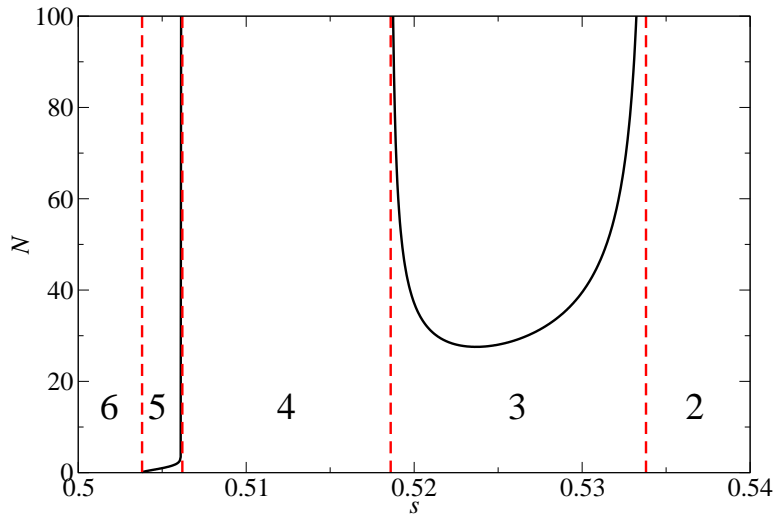}}
\caption{The total number of $e$-foldings versus the value of $s$.
Dashed lines and numbers identify the regions described in the text.}
\label{Ns_c}
\end{figure}

We now consider how the inflationary potential and the resulting
solutions depend on $s$.  
\Fig{Ns_c} shows the total number of $e$-foldings as a
function of $s$.
The parameter space of $s$ can be divided into six regions. They are\footnote{In the flat region of the potential where
inflation takes place, the Coulomb term is typically unimportant;
therefore the correspondence between $D(x)$ used by previous authors
and us is approximately $D(x) \to D_1(1 + D_{01})$; hence we would take
$D_1$ to be smaller by a factor of $(1 + D_{01})$ to reproduce their
results.  Moreover, since the second term
gives a vanishing contribution to uplifting at $x=0$, the
correspondence between previous authors' value of $s$ and ours is
$s\to s/(1 + D_{01})$.  This explains why we need $s$ near $0.5$
for getting a flat potential, while Refs.\ \cite{Baumann, Panda}
had $s\sim 1$ for the same parameters.}

\begin{enumerate}
\item $s> 3.443$: $dV/d\omega \not= 0$; there is no valley-shaped
 potential and no stable trajectory.
\item $0.5338<s\le3.443$: the inflaton gets trapped in a local minimum, 
$N\to\infty$.
\item $0.5186\le s\le0.5338$: the potential is monotonic; 
for $s\to0.5186$ or 0.5338, it is extremely flat and gives a large 
number of $e$-foldings.
\item $0.5062\le s <0.5186$: the inflaton is again trapped
 in a local minimum, so $N\to\infty$.
\item $0.5038\le s <0.5062$: there is a local minimum in the
potential;  however the inflaton has enough momentum to escape 
from it if slow roll had been attained earlier 
(e.g., starting with zero initial velocities $\dot x_i= \dot 
\omega_i=0$ at $x_i=0.8$).
\item $0<s<0.5038$: the potential is negative at $0<x<0.8$, so
inflation ends in a big crunch.
\end{enumerate}

We are interested in the cases where inflation will end (although not
via a big crunch). Although region 5 allows for the inflaton to
escape from the local minimum and end inflation, \fig{Ns_c}
shows that one needs to extremely fine-tune the value of $s$ to get
enough $e$-foldings. Therefore we will focus on region 3, where the
potential is monotonic, and much less tuning is needed.  It is
striking that there is a plateau in this region, signifying  a {\it
minimum} number of $e$-foldings of about 30. It is tempting to look for
other parameters such that the minimum number of $e$-foldings of the
plateau would be increased to more than 50.  If that were possible, 
then one would say that inflation is generic rather than fine-tuned,
at least with respect to the parameter $s$.  To investigate this
possibility, we need to search the multidimensional parameter space.
This will be discussed in Section \ref{metro}.

Region 3 is also the case that Ref.\ \cite{Panda} focused on. The
latter found that by adjusting the tension, one can have inflation
with a correct spectral index and enough $e$-foldings, but they did not
succeed in finding a model which satisfies the COBE normalization
simultaneously.  We confirm this: by restricting to the default
parameters in Ref.\ \cite{Baumann}  and just varying the value of
$s$, one cannot achieve  $n_s\sim 0.96$ and ${\cal P_R} \sim
2.4\times10^{-9}$ at $N\ge50$.

\subsection{Varying $A_0$: COBE normalization}
\label{varyingA0}

Of course, by varying just one parameter, one should not expect to
satisfy several experimental constraints.  One usually realizes the
COBE normalization by adjusting the overall scale of the potential.
In Ref.\ \cite{Baumann}, it was assumed that the prefactor $A_0$
played the role of the overall scale.  However this is only true
as long as the Coulombic interaction is negligible, because as
shown in \eq{D1}, $A_0$ is related to $D_1$ and hence contributes to
the shape of the Coulomb interaction through \eq{VD}.\footnote{Even
if we Taylor expand the Coulomb term as is done in Refs.\
\cite{Baumann:2007np, Baumann, Panda}, the shape of the Coulomb
interaction is still related to $A_0$.} Therefore as long
as $D_{01}>0$, which must be the case, $A_0$ does not
merely determine the overall scale of the potential.

\begin{figure}[htp]
\centering{\includegraphics[width=4in]{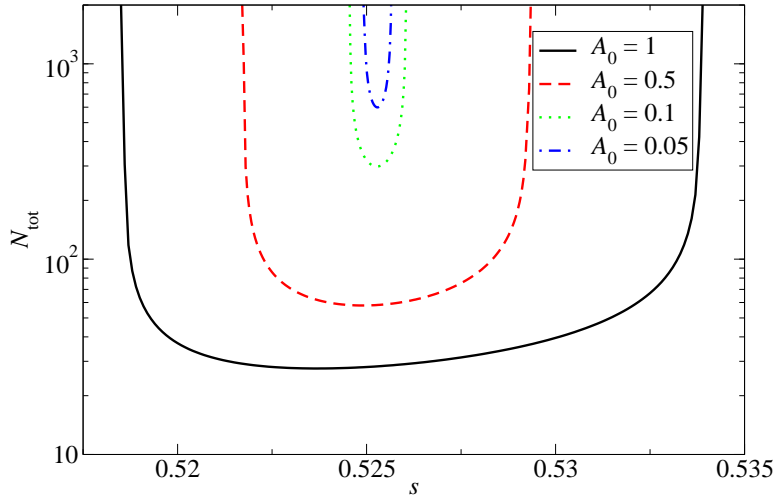}}
\caption{The total number of $e$-foldings versus the value of $s$ 
with different values of $A_0$.}
\label{Ns_A0_c}
\end{figure}

This can be seen explicitly by calculating the total number of
$e$-foldings of inflation versus $s$ for varying values of $A_0$. If
the shape of the potential did not depend on $A_0$, then the number
of $e$-foldings would be completely insensitive to $A_0$, as long as
the slow-roll approximation is valid. \Fig{Ns_A0_c} shows that in
fact $N_{\rm tot}$ depends rather strongly on $A_0$:  for smaller
values of $A_0$ (closer to those needed for the COBE normalization),
the range of $s$ values which give a monotonic potential becomes
smaller. An interesting by-product is that the minimum number of
$e$-foldings becomes bigger. This shows that when $A_0$ is adjusted to
satisfy the COBE normalization, $s$ does not need to be fine-tuned to
a special value  to ensure that the number of $e$-foldings of inflation
is sufficient. However, $s$ still needs to be within a narrow range
to avoid the problem of the potential developing a local minimum in
which the inflaton gets stuck, with no end to inflation. For example,
taking $A_0 = 0.005$ and $s=0.5253$, we have ${\cal P_R} =
2.41\times10^{-9}$ at $N = 73$ ($N_{\rm tot} \simeq 6000$), where
$n_s = 0.944$, within 2$\sigma$ of WMAP5's result
\cite{Komatsu:2008hk}.\footnote{\label{fn4}The marginalized values
(mean and 95\% C.L.) from WMAP5 are ${\cal P_R} =
(2.41\pm0.22)\times10^{-9}$ and $n_s = 0.963\pm0.028$.} We have an 
existence proof that it is possible to nearly satisfy all the cosmological constraints
in the $D3$-$\overline{D3}$ inflation model; however this particular
example appears to be fine-tuned.  Below we will show that less
fine-tuned examples can be found.

\subsection{Accounting for the uplifting constraint}
\label{uplifting}

Up to now we have regarded $s$ as a free parameter,  as
was done in the previous literature \cite{Baumann:2007np, Baumann,
Panda}. But as we have mentioned, $s$ should already be determined by
the requirement of having vanishing cosmological constant at the end
of inflation.  In this section, we take this requirement into
account and show that it is nevertheless still possible to tune to
obtain a flat
potential using the extra parameter $D_{01}$, which in fact has a
similar qualitative effect to varying $s$.

First we show how the value of $s$ is fixed by setting
\be
  V(0,\omega_0) = 0.
\ee
Using \eq{w0}, $\omega_0=\omega_0(s,\omega_F)$, we find that
\be
  (2\omega_F+3)e^{\omega_0-\omega_F}-2\omega_0-5=0.
\label{wF}
\ee
This shows that $\omega_0$ can be expressed 
as a function of $\omega_F$ only, $\omega_0=\omega_0(\omega_F)$. 
From \eqs{w0}{wF}, we have
\bea
  s &=& \frac{\omega_0+2}{\omega_F}\left(\frac{2\omega_F+3}{2\omega_0+5}\right)^2\\
  &\simeq&{\omega_F\over\omega_0}\left[1+3\left({1\over\omega_F}-{1\over\omega_0}\right)\right],
\eea
where the latter form assumes $\omega_0$, $\omega_F\gg 1$.\footnote{In most cases, $\omega_0\simeq\omega_F,$ so $s\sim1$.} Therefore,
once $\omega_F$ is given, both $\omega_0$ and $s$ are fixed by
uplifting $V(0,\omega_0)=0$, and our use of $s$ to flatten the
potential in the previous sections is seen to be invalid. 
For example, taking the value of $\omega_F$ 
in \eq{fid2}, we obtain $\omega_0 = 10.100,\ s = 1.0087$.

\begin{figure}[htp]
\centering{\includegraphics[width=4in]{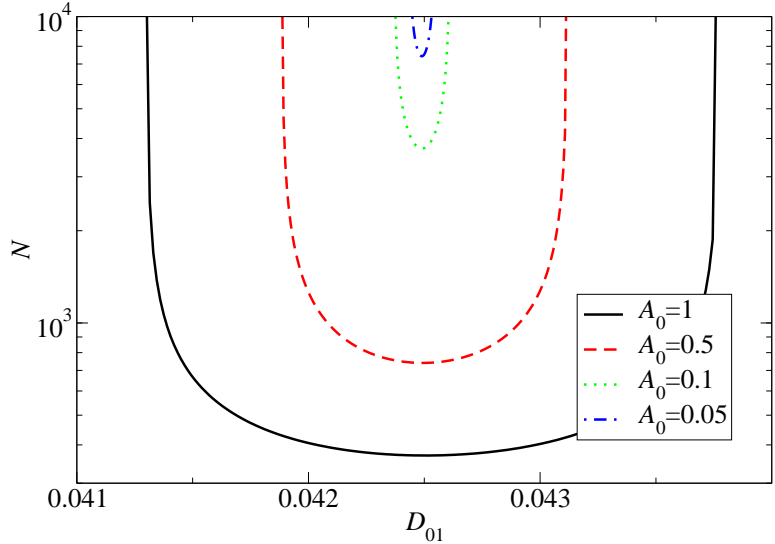}}
\caption{The total number of $e$-foldings versus the value of $D_{01}$ with different values of $A_0$.}
\label{N_D01}
\end{figure}

To compensate for not being able to vary $s$, 
we can adjust the ratio
of the tensions in the inflationary versus the other throats, i.e., $D_{01}$, while keeping $s$ fixed.   We find that the
qualitative dependence of the shape of the potential on $D_{01}$ is
similar to the dependence on $s$.  
There are five regions of the  $D_{01}$ parameter space which
correspond to the first five enumerated for $s$ in Section
\ref{varyings}.  The sixth region, where the potential became
negative, no longer exists because we have now adjusted $s$ to avoid
this problem.  Furthermore, the dependence of the shape of the curves
$N_{\rm tot}(D_{01})$ on the parameter $A_0$ is just like that for
$N_{\rm tot}(s)$, as can be seen in \fig{N_D01}.  Again, as $A_0$
decreases, the minimum number of $e$-foldings of the plateau increases,
while the monotonic region shrinks and more fine-tuning is needed.

We find that varying only $(D_{01},A_0)$ while keeping other
parameters fixed is not sufficient for satisfying all the
experimental constraints, but varying $(D_{01},\omega_F)$ allows
us to nearly do so.  For example, taking $\omega_F=15$ and $D_{01}=0.1227$ gives
${\cal P_R}=2.41\times10^{-9}$ and $n_s=0.949$ at $N=78$, out of  the
total number of $e$-foldings $N_{\rm tot}\simeq10^4$.  Generically
we might have found that by varying two parameters, the two
constraints ${\cal P_R}\sim 2.4\times10^{-9}$ and  $n_s \sim 0.96$
could be satisfied at the COBE scale, which we take to be $50\le N_e\le60$.
The fact that we cannot do so here is just due to unfortuitous values
for some of the other parameters which are not varied.  
In the next section
we will consider variations in the full parameter space, but  
here we  are primarily concerned with  the degree of fine-tuning
needed to get sufficient inflation.  The present example 
requires fine-tuning at the level of $0.13\% 
$ for $D_{01}$:
the potential is monotonic from $D_{01}=0.122\,62$ to
$0.122\,78$, so the relative fine-tuning is
$1.6\times10^{-4}/0.1227$, approximately 1 part in 1000.
Below, we will identify other regions of parameter
space where this problem is significantly alleviated.

\section{Solving the Fine-Tuning Problem}
\label{metro}

Although the required values in the previous section may
subjectively appear to be rather finely tuned, this is a notion which
requires definition.  One must distinguish fine-tuning from the
more mundane necessity of fixing parameters to agree with
experimental measurements.  In this section we propose a measure on
the volume of the experimentally allowed part of parameter space
which will allow us to quantify the degree of fine-tuning needed in
any localized region of the space.  We calculate this statistic while
doing a systematic search of the parameters.  The results are
described in the final part of this section, demonstrating that the
fine-tuning problem is ameliorated for optimal values of the
parameters.

Our goal now is to 
scan the parameter space in search of regions where less
tuning is required.   To make this quantitative, we
need some specific measure of the degree of tuning, which varies locally
in the parameter space.  Suppose we have identified a set of
parameters $p_i^{(0)}$ (where $i$ runs over the number of parameters)
which satisfy some necessary criteria for inflation.  
We can then vary each parameter to find the maximum range
$p_i^{(0)}-\sigma_i < p_i < p_i^{(0)}+\sigma_i$ for which these
criteria are still satisfied.\footnote{In general the interval will 
not be symmetric; it 
has the form $p_i^{(0)}-\sigma_i^- < p_i < p_i^{(0)}+\sigma_i^+$.
To simplify computations, we take $\sigma_i$ to be
the minimum of these two values, which underestimates the allowed
volume.}  The width of this interval 
is $2\sigma_i$.

A first guess for a quantity which is anticorrelated with the degree
of fine-tuning would be the volume in parameter space consistent with
inflation, defined by the product of all the intervals $2\sigma_i$. 
The exact $N$-dimensional volume would be more complicated than 
our rectilinear approximation; we ignore this, and treat the $\sigma_i$'s
as independent quantities, i.e., when determining
 $\sigma_i$, we fix other parameters at
their central values, $p_j^{(0)}\ (i\not=j)$.  The volume is  thus given by
\be
  V_N\left(p_i^{(0)}\right) = 2^N\prod_i \sigma_i.
\label{Vneq}
\ee
However it would be naive to think that maximizing \eq{Vneq}
corresponds to minimizing the tuning of parameters, because this
would artificially reward parameter values that happen to be
large in absolute terms.\footnote{On the other hand, it can also be argued that 
the absolute volume is also a reasonable measure of the tuning,
and we will come back to consider it in  
Section \ref{measure}; there we will show that our conclusions are not
sensitive to this choice.} We are really interested in
the relative variation of a given parameter.  Therefore a better measure
of naturalness is the relative volume,

\be
  \delta_N = \prod_i {2\sigma_i \over p_i^{(0)}} = 
{V_N \over \prod\limits_ip_i^{(0)}}.
\label{Vrel}
\ee
We can furthermore define a reduced relative volume,
\be
  \delta_N^\prime = \sqrt[N]{\delta_N},
\ee
whose reciprocal quantifies the average degree of tuning per
parameter, and thus makes it meaningful to compare searches in which
different numbers of parameters are varied.

\subsection{Description of algorithm}

To explore the parameter space, we used the Metropolis algorithm,
which looks for a function's minimum  by the method of simulated
annealing \cite{nr}. In our case, we chose the objective function 
$f_{\rm  obj}$ to be the negative of the relative volume, \eq{Vrel}. Because we are interested in the part of parameter space
corresponding to \figs{Ns_c}{Ns_A0_c}, where there is
always a minimum number of $e$-foldings of inflation, we chose as our
criterion for successful inflation that the potential should be
monotonic, rather than having a local minimum where inflation would
never end.  In addition, we want to select models that satisfy the
experimental constraints.  We do this by requiring that the central
values $p_i^{(0)}$ of the regions correspond to experimentally 
allowed models, although
we do not impose this additional requirement on the neighboring
points that define the volume. 
Specifically, for the central point of each allowed
volume we demand that the COBE normalization (${\cal
P_R}=2.41\times10^{-9}$) is satisfied with the number of $e$-foldings $50 \le N_e \le 60$ before
the end of inflation.  We take this to be a reasonable reflection of
uncertainties in the time of horizon crossing due to  variations in
the scale of inflation and reheat temperature.  We also
require the spectral index to be within 2$\sigma$ of the WMAP5
preferred value, $n_s = 0.963\pm0.028$ \cite{Komatsu:2008hk}.\footnote{The power spectrum is insensitive to the location of the
inflaton. For example, if one parametrization gives $N_e=55$
and $n_s=0.963$ at ${\cal P_R}=2.41\times10^{-9}$, then this
parametrization (with a different inflaton location) will roughly
give the same $N_e$ and $n_s$ at ${\cal
P_R}=(2.41\pm0.22)\times10^{-9}$ (WMAP5's 2$\sigma$). Therefore we
fix the normalization at WMAP's mean value for simplicity.}

The motivation for measuring volumes which are consistent with
inflation but not necessarily the particular experimental constraints
we observe is the following.  The volume of parameters satisfying
some number of constraints would be a set of measure zero in the full
parameter space.  However we do not consider a model to be fine-tuned
just because its parameters are fixed by certain measurements. 
Rather, it is the basic requirement of having a flat enough potential
to get at least 60 $e$-foldings of inflation (and not getting stuck in a
local minimum preventing an exit from inflation) which underlies the
apparent need for tuning that we are interested in.

The actual objective function includes a few subtleties. For example,
if a set of parameters does not give a monotonic potential nor
satisfy the experimental constraints, then the relative volume can be
defined to be zero.  
However, this gives no information to assist the Monte Carlo method in
finding more favorable values since most points in the parameter space will have the same value of the objective function. In this case we therefore take the
objective function (which is to be minimized) to be an
empirical function of $N_e$, $\cal P_R$, and $n_s$:
\be
  f_{\rm obj} = \left\{
  \begin{array}{ll}
    f_{\rm emp}(N_e, {\cal P_R}, n_s),&\hskip0.5in\delta_N=0,\\
    -\delta_N,&\hskip0.5in\delta_N>0.
  \end{array}\right.
\ee

The empirical function is chosen in such a way as to
help move the configuration from a nonmonotonic regime to a monotonic
one, or from a regime not satisfying the experimental constraints to
one which does. Once satisfactory parameters are found, then the
relative volume is calculated, and maximizing $\delta_N$ leads to
parameter values which are less fine-tuned.\footnote{The
empirical function is not essential; it accelerates the search, 
but it has no effect on the results once a
monotonic regime satisfying the constraints is found. The particular
function used in this paper is given  in Appendix \ref{empirical},
but one is free to design a different one.}

The Metropolis algorithm uses an artificial temperature $T$ which randomly allows
the objective function to sometimes increase rather than decrease; this is how it avoids
getting stuck in a shallow local minimum rather than converging to
some point closer to the global minimum.   We used the downhill
simplex method of Ref.\ \cite{nr} as a generator of random steps,
which moves the system's configuration by reflections, expansions,
and contractions in an $N$-dimensional simplex.  As the temperature
is lowered, the system relaxes to a minimum which should be close to
the global minimum.  During this iterative process, a chain of
accepted parameter values is generated, which allows one to make
statistical statements about the probabilities of the parameters.

\subsection{Monte Carlo results}

Our starting point was the configuration given in the previous
section, with the values
\be
D_{01} = 0.122\,70,\ \omega_F = 15,\ A_0 = 1,\ \phi_\mu = 0.240\,56,\
n = 8.
\label{fid}
\ee
Since we had not yet varied enough parameters, this does not quite
satisfy all the experimental constraints: it has spectral index and
normalization $n_s = 0.949$ and ${\cal P_R} = 2.41\times10^{-9}$ at
$N=78$ $e$-foldings before the end of inflation, instead of at the COBE
scale. Nevertheless, since it is closer to the examples previously
studied in the literature, we will use this as a reference point for
comparison when assessing the improvement in fine-tuning. We also
restarted the search using other initial conditions  to avoid  getting trapped
in a local minimum.   We accumulated approximately 200 chains
containing more than 70\,000 samples (each chain containing from 200
to 1000 samples).  The control parameter $T$ was decreased by a
factor $\epsilon$ after every $m$ moves, over several orders of
magnitude (for example, changing from  100 to 0.001)  during the
entire simulated annealing process. The computing time for each chain
can be as fast as five hours (on a PC), or as slow as ten days,
depending on the configuration of the annealing schedule and the
initial conditions.

To satisfy  the experimental constraints  we tried varying different
combinations of the five independent parameters, $(D_{01}, \omega_F,
A_0, \phi_\mu, n)$.   Generically one would expect that any two
parameters would be uniquely fixed by the two constraints; thus to
obtain chains, one should vary at least three at a time.  Indeed, we
found that the combinations  $(D_{01},\omega_F,A_0)$,
$(D_{01},\omega_F,\phi_\mu)$, or $(D_{01},\omega_F,n)$ were suitable
for generating chains which evolved toward larger reduced volumes.
In each case, the unvaried parameters take the values given in
(\ref{fid}), which we refer to as the starting or fiducial point.
The results are shown in Table \ref{tab1}, where it can be seen that
the reduced volumes of parameter space grow to values of order unity,
starting from very small initial values.  This indicates that there
is essentially no fine-tuning at the optimal points.

\begin{table}[htp]
\caption{The relative volumes and reduced volumes
for different combinations of the parameters, showing initial and
final values along the chains.}
\label{tab1}
\renewcommand{\arraystretch}{1.5}
\vspace{2ex}
\centering{\begin{tabular}{|c|cc|cc|}
\hline\hline
\raisebox{-1.7ex}[0pt][0pt]{Parameters} & \multicolumn{2}{c|}{Starting point} & \multicolumn{2}{c|}{Optimal point}\\
& $\delta_3$ & $\delta_3^\prime$ & $\delta_3$ & $\delta_3^\prime$\\
\hline
$(D_{01},\omega_F,A_0)$ & $9.0\times10^{-7}$ & 0.96\% &\ 0.087 & 44\%\\
$(D_{01},\omega_F,\phi_\mu)$ & $5.2\times10^{-10}$ & \ 0.080\% & 0.027 & 30\%\\
$(D_{01},\omega_F,n)$ & $7.8\times10^{-10}$ & 0.092\% & 0.020 & 27\%\\
\hline\hline
\end{tabular}}
\end{table}

\begin{figure}[htp]
\centerline{\includegraphics[width=0.5\hsize]{\figdir wF-A0.epsi}
\includegraphics[width=0.5\hsize]{\figdir 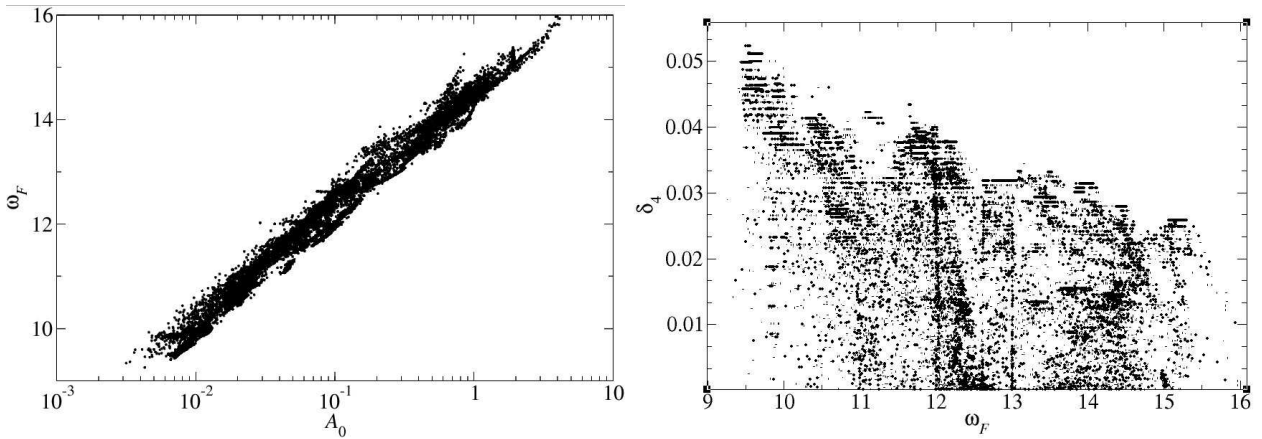}}
\caption{Left panel: scatter plot of accepted  $\omega_F$ and $A_0$ values
from search of 4D parameter space $(D_{01},\omega_F,A_0,\phi_\mu)$.
Right panel: correlation of $\delta_4$ and $\omega_F$.}
\label{wF-A0}
\end{figure}

We can make more detailed statements about how the various parameters
affect the power spectrum.  For example, increasing $\omega_F$ has
the effect of reducing the overall scale of the potential, due to
the $e^{-2\omega}$ dependence, while increasing $A_0$ has the opposite
effect.  It is not surprising that the COBE normalization thus
produces a strong correlation between $A_0$ and $\omega_F$ for the
accepted parameter values, which can be seen in \fig{wF-A0} (left panel).\footnote{We note that the first entry in Table
\ref{tab1} ($\delta_3=0.087$) for
$(D_{01},\omega_F,A_0)$ is unphysical because we did not apply the
constraint $A_0<1$ (to avoid super-Planckian gaugino condensate scales)
there.   Starting from the fiducial point (\ref{fid}), one cannot
increase $\delta_3$ by varying only ($D_{01},\omega_F,A_0$) while 
requiring $A_0 \le 1$.
However the 4D parameter space search described below
allowed us to find nontuned examples with $A_0<1$.} Also shown in
that \fig{wF-A0} (right panel) is the correlation of 4D relative volumes
with the value of $\omega_F$ for the Monte Carlo chains.  The latter
demonstrates that larger values of $\omega_F$ are less likely than
smaller ones, but
only mildly so.

Furthermore, we find a  degeneracy between $\phi_\mu$ and $n$ on the
shape of the potential: the effect of the discrete parameter $n$ (the
number of D7-branes in the stack) can be compensated by changing the
value of $\phi_\mu$.  Therefore the subset
$(D_{01},\omega_F,A_0,\phi_\mu)$ gives an adequate representation of
the possible potential shapes arising in the model.  Fixing $n=8$ to
eliminate the $\phi_\mu$-$n$ degeneracy,  the Metropolis algorithm
finds a global minimum of $f_{\rm obj}$, hence a maximum of the
relative volume $\delta_4=0.052$. Table \ref{tab2} compares the
optimal point to the fiducial point given in the previous section for the
4D search.  We see that the fiducial point required tuning at the
level of $0.5$\% per parameter; this number increases to 50\% at the
optimal point, so there is no more tuning.  We also show the
breakdown on a per-parameter basis: $2\sigma_i/p_i^{(0)}$ is the
relative allowed width for the $i$th parameter.  One might worry that
the cruder statistic $\delta'_4$ could hide severe tuning of some
parameters by having very large values of $2\sigma_i/p_i^{(0)}$ for
others, to which inflation happened to be insensitive.  However we
see from Table \ref{tab2} that this is not the case: the most
sensitive parameter is $D_{01}$, which is still only tuned at the 
20\% level, in the region of the optimal parameter values.
The last row of the table shows
that the tuning per parameter is ameliorated by at least a factor
of 100 (except for $A_0$, which did not require fine-tuning even
at the starting point).\footnote{Recall that we impose three experimental
constraints (${\cal P_R}, n_s, N_e$) on the central point
($p_i^{(0)}$) of the volume, but we only require a monotonic
potential for points in the volume,
$p_i^{(0)}-\sigma_i<p_i<p_i^{(0)}+\sigma_i$. To solve the horizon
problem, we need $N_{\rm tot}\gsim60$; adding this requirement to the
volume, the $2\sigma_i/p_i^{(0)}$ for $\phi_\mu$ in the table will be
changed from 0.3 to 0.2. However, the modifications to the total
volumes are small, $\delta_4 = 0.036$ and $\delta_4^\prime = 44\%$.
And the conclusion remains true.}

\begin{table}[htp]
\caption{Comparison of parameter values and degree of fine-tuning
between the fiducial (starting) point and the optimal point of 4D parameter
search.}
\label{tab2}
\renewcommand{\arraystretch}{1.5}
\vspace{2ex}
\centering{\begin{tabular}{ccccccc}
\hline\hline
Configuration & $D_{01}$ & $\omega_F$ & $A_0$ & $\phi_\mu$ & $\delta_4$ & $\delta_4^\prime$\\
\hline
Fiducial point & 0.122\,70 & 15 & 1 & 0.240\,56 & $4.5\times10^{-10}$ & 0.46\%\\
$2\sigma_i/p_i^{(0)}$ & $1.3\times10^{-3}$ & $\ 8\times10^{-4}$ & 0.86 & $5\times10^{-4}$ & -- & --\\
\hline
Optimal point & 0.1976 & 9.550 & 0.007778 & 0.5894 & 0.052 & 48\%\\
$2\sigma_i/p_i^{(0)}$ & 0.19 & 0.52 & 1.8 & 0.3 & -- & --\\
\hline
$\left[\sigma_i/p_i^{(0)}\right]_{\rm opt}$/$\left[\sigma_i/p_i^{(0)}\right]_{\rm fid}$ & 140  & 640 & 2.1 & 600 & -- & --\\
\hline\hline
\end{tabular}}
\end{table}

\begin{figure}[htp]
\centering{\includegraphics[width=0.5\hsize]{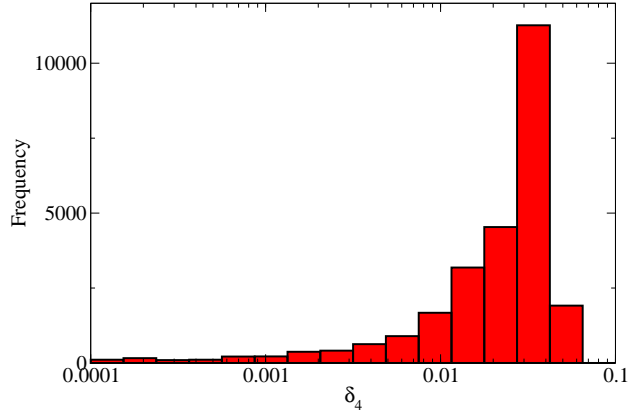}}
\caption{Distribution of relative volume $\delta_4$ from
the Monte Carlo chains.  Notice the scale for $\delta_4$ is
logarithmic.}
\label{histd4}
\end{figure}

\begin{figure}[htp]
\centering{\includegraphics[width=\hsize]{\figdir 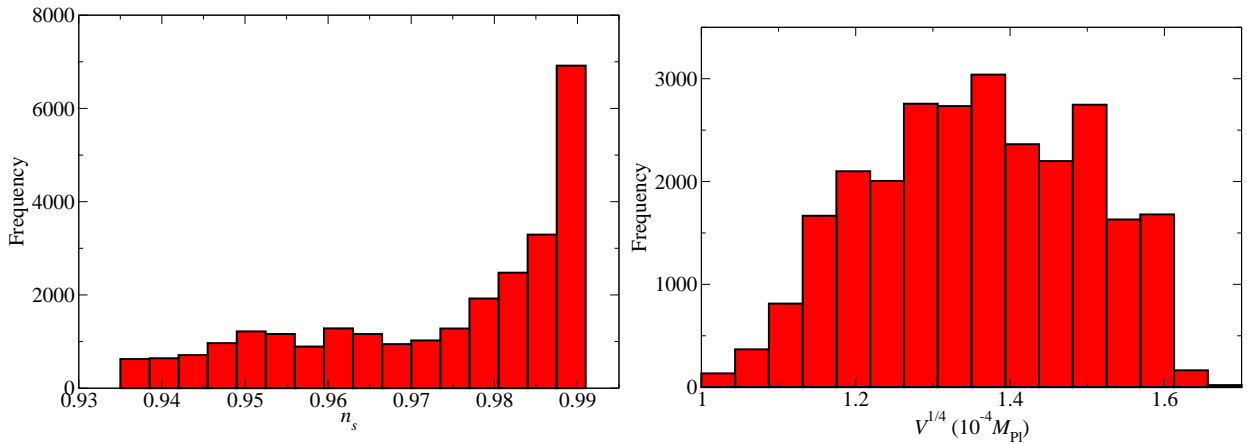}}
\caption{Distribution of spectral index $n_s$ and inflationary
scale, based on Monte Carlo chains.}
\label{hist}
\end{figure}

Given that points in our chains tend to accumulate where the relative
volume is bigger and the tuning problem is less severe, we can use
the chains to define a probability distribution on the space of
parameters, as well as on the predictions of the model for observable
quantities.  The correlation between relatively large volume and high
density of models can be seen in the distribution of $\delta_4$,
\fig{histd4}.  
The distributions for the spectral index (within the
$2\sigma$ range which we allowed around the WMAP5 central value) and
the energy scale of inflation are shown in \fig{hist}.  We
note there a preference for larger values of $n_s$ close to $0.99$,
while $V_{\rm inf}^{1/4}$ is in the range $(1.1$-$1.6)\times 10^{-4} M_{\rm Pl}
= (2.7$-$3.9)\times 10^{15}$ GeV.\footnote{$M_{\rm Pl}$ is the reduced Planck
mass, $2.44\times 10^{18}$ GeV.}

\subsection{An alternative measure}
\label{measure}

Thus far, we have considered a well-defined and reasonable definition
for the measure of fine-tuning on the parameter space: given a point
which is consistent with the data, it measures the relative amount by
which each parameter can be varied around this point and still be
acceptable.  However, one might argue that this hides fine-tuning
problems in the case where the central values are orders of magnitude
smaller than the largest theoretically allowed values (similar to
the  cosmological constant problem).  In this section, we therefore
consider an alternative measure, which is the absolute volume of
parameter space consistent with measurements, \eq{Vneq}, instead of the relative
volume.

The absolute parameter volume as a measure of fine-tuning has
some conceptual difficulties which are not present in the relative
volume measure.  First, it is possible that some parameters have a
formally infinite range, such as the K\"ahler modulus $\omega_F$. 
One has to regularize this divergence in order to make any sensible
statements about probability.  Second, we must disentangle the issue
of fixing parameter values through measurements from that of
fine-tuning.  For example, in the theory of quantum electrodynamics,
we do not say that the charge of the electron is fine-tuned just
because its value is experimentally known to a very high precision.
The tuning problem we are concerned with is one where different
free parameters have to be adjusted with respect to each other to
a high precision in some artificial way that had no {\it a priori}
justification.  Thus to implement the absolute volume measure, it
makes sense to consider the volume only of some linear combinations
of parameters which are not fixed by experimental constraints.  This
has the added advantage of possibly solving the first problem: we can
use the experimental constraints to fix values of parameters which 
have no theoretical upper bound.   

There is a further shortcoming in the absolute volume measure when
degeneracies exist between parameters. In the present case, the
scale of  the inflaton potential is determined by the product of two
prefactors,  $A_0^2$ and $e^{-2\omega_F}$. As a result, a
degeneracy between $\ln A_0$ and $\omega_F$ exists, as is  confirmed
by \fig{wF-A0} (left panel). One can achieve a large absolute volume by
trading $\omega_F$ for $A_0$. For instance, we find models with
super-Planckian values of $A_0$ of order $O(10^6)$
($\omega_F\simeq30$) having $\sigma_{A_0}\sim O(10^6)$ and $V_4\sim
O(10^4)$,\footnote{$A_0>1$ is theoretically disfavored; here we just 
use it to illustrate the effect of the degeneracy.} whereas  the
absolute volume near the fiducial point is $2\times 10^{-10}$; see
Table \ref{tab3}. But it would be misleading to claim that the
model $(\omega_F,A_0)\sim(30,10^6)$ has less fine-tuning than that
with $(\omega_F,A_0)\sim(10,0.01)$, because the large $V_4$ is a
result of the fact that the potential has different sensitivities to
$\omega_F$ and $A_0$. The relative volume, on the other hand, can
balance the different sensitivities between different parameters.

Despite these complications, it might still be of interest
to know the ratio of the allowed volume to the total volume of the
parameter space,
\be
  \label{deltaNmax}
  \Delta_N = \frac{V_N}{\prod\limits_ip_{i,{\rm max}}},
\label{volratio}
\ee
where $p_{i,{\rm max}}$ denotes the theoretical maximum of the
parameter.  As mentioned above, $\omega_F$ has no maximum value,
and the same is true of $D_{01,{\rm max}}$.  Before dealing with this,
let us consider just the numerator in (\ref{volratio}), $V_N$ itself.
We have repeated the Monte Carlo search of parameter space to 
find regions where $V_N$ are maximized.  The results are shown in 
Table \ref{tab3}.
There it can be seen that while $V_N$ is only 
$2\times10^{-10}$ for the fiducial (starting) point, it increases
to $0.03$ at the optimal point, which is different from the optimal
point for the relative volume shown in Table \ref{tab2}. 
It is interesting that the improvement, 8 orders of magnitude, 
is the same as we obtained using the 
relative volume as the measure. Our conclusion that the degree of
delicateness depends on the parameter values, and is greatly
ameliorated in some regions of parameter space compared to others,
thus holds for both the relative volume measure and
the absolute volume measure.

\begin{table}[htp]
\caption{Comparison of parameter values and absolute allowed volume between the fiducial (starting) point and the optimal point of 4D parameter search.}
\label{tab3}
\renewcommand{\arraystretch}{1.5}
\vspace{2ex}
\centering{\begin{tabular}{cccccc}
\hline\hline
Configuration & $D_{01}$ & $\omega_F$ & $A_0$ & $\phi_\mu$ & $V_4$\\
\hline
Fiducial point & 0.122\,70 & 15 & 1 & 0.240\,56 & $2\times10^{-10}$\\
$2\sigma_i$ & $1.6\times10^{-4}$ & 0.012 & 0.86 & $1.2\times10^{-4}$ & -- \\
\hline
Optimal point & 0.2186 & 13.70 & 0.6341 & 0.4383 & 0.03\\
$2\sigma_i$ & 0.06 & 11 & 0.73 & 0.063 & --\\
\hline
$\left[\sigma_i\right]_{\rm opt}$/$\left[\sigma_i\right]_{\rm fid}$ & 380  & 920 & 0.85 & 530 & --\\
\hline\hline
\end{tabular}}
\end{table}

Although the optimal case is much less fine-tuned than the fiducial
point, we are still left with the question of how fine-tuned is it in
an absolute sense, which would be answered if $\Delta_N$ was
well defined.  If we adopt the approach suggested above, of
eliminating the parameters which have no maximum value ($\omega_F$
and $D_{01}$), then we are left with $\Delta_2$ for the remaining
parameters $A_0$ and $\phi_\mu$.  The maximum value of 
$\phi_\mu$ is 1, the maximum initial
separation between the D3-branes and anti-D3-branes; similarly, $A_0$
should be less than the Planck scale due to gaugino condensation.
This gives $\Delta_2 = 0.046$ and $\sqrt{\Delta_2} = 0.2$ for the mean
degree of tuning per parameter.  This is comparable to the result we
obtained using the relative volume measure.  

Alternatively, instead of eliminating the $D_{01}$ and $\omega_F$
parameters, we could compare their widths to some ``reasonable'' or
``generic'' values.  Since  $D_{01}$ is the ratio of two brane
tensions, $D_{01}\sim 1$ would seem to be a generic value.  As for
$\omega_F$, even though in principle it could be arbitrarily large,
in practice it is difficult to make it very large.  It is a derived
parameter, which  depends upon $W_0$ through \eq{W0}. For example, if
$W_{0,{\rm min}}\sim10^{-42}$, then $\omega_{F,{\rm max}}\sim 100$.  
We see that in fact it requires a high degree of fine-tuning of $W_0$
to make $\omega_{F}$ very large.  Generously taking 100 to be its
maximum value (even though this would be considered unreasonably large
by most string theorists), we then find that $\Delta_4 = 3\times 10^{-4}$,
$\sqrt[4]\Delta_4 \sim 0.1$, again a quite modest level of tuning per
parameter.  

\section{Properties of the Optimal Parameter Set}
\label{discussion}

Having identified a favorable region in the space of the model
parameters,\footnote{We use the optimal point given in Table \ref{tab2} in this section, but the results also hold for the optimal point in Table \ref{tab3}.} we now consider a number of its detailed properties,
including its consistency with string theoretic constraints, the
shape of the inflationary trajectory, and sensitivity to initial
conditions and the overshoot problem.  We also explain a
potential subtlety concerning the computation of the spectral index
in the single-field approximation to the model.  

\subsection{Microscopic parameter values}

The parameters ($D_{01}, \omega_F, A_0, \phi_\mu$) which were convenient
to vary in the potential are not all fundamental from the 
string theoretic point of view. 
We would like to determine the values of 
the microscopic stringy parameters which are compatible with
the optimal point  in Table \ref{tab2}, which has
\be
  D_{01} = 0.1976,\quad \omega_F = 9.550,\quad A_0 = 0.007\,778,\quad \phi_\mu = 0.5894, \quad n = 8.
\label{optimal}
\ee
These yield the derived parameters $\omega_0 = 9.644$, $s = 1.0094$, and
the corresponding observational predictions are 
${\cal P_R} = 2.41\times10^{-9}$ for the primordial power spectrum and 
$n_s = 0.989$ for the spectral index, at $N=52$ $e$-foldings before the
end of inflation, out of a total of $N_{\rm tot} = 134$ $e$-foldings
of inflation.
It is straightforward to find a reasonable set of stringy parameters
$(N_5,B_4,B_6,Q_\mu)$ giving the desired $(\phi_\mu,\omega_F)$. For
example, taking $Q_\mu=1.07$ and $N_5=10$, we need $B_4 = 75.28$,
$B_6 = 1.0058$.  This set satisfies all the 
microscopic consistency conditions in Appendix \ref{microscopic}.

One might however be concerned that such a small value of $\omega_F$
as 9.6 is only marginally consistent with the need for control over
higher derivative corrections to the low-energy effective action,
which are suppressed by the large compactification volume (hence
small curvatures).  \Fig{wF-A0} shows that larger values of 
$\omega_F$ are indeed possible, up to a maximum of $\omega_F\simeq
15$; beyond this point, overly large values of
$A_0>1$ would have to compensate the reduction in the inflationary
scale needed to get the right COBE normalization.  Thus one can
increase $\omega_F$ to somewhat larger values, but at the expense of
saturating the consistency condition $A_0<1$, and somewhat increasing
the degree of  fine-tuning. 
However at $\omega_F\simeq 14.5$, where the constraint $A_0<1$ starts
to become important, 
there is actually no fine-tuning: the reduced volume becomes 
$\delta_4\simeq 0.021$ at this
point, as opposed to  the optimal value $\delta_4\simeq 0.052$.  (Recall
that  $\delta_4^\prime=\sqrt[4]{\delta_4} = 38$\% quantifies the degree of tuning per
parameter.)

\subsection{Taylor expansion of DBI kinetic term}

Another point of consistency concerns the expansion of the DBI action
for the inflaton kinetic term  $-a^4 T_3\sqrt{1-a^{-4}\dot X^2}$
where $T_3 = m_s^4/(8\pi^3 g_s)$ is the 3-brane tension and $a$ is
the warp factor in the throat. Notice that the canonically normalized
inflaton field is $\phi\sim \sqrt{T_3} X$ after Taylor expanding this
expression.  Estimating  the inflation scale as $V_{\rm
inf}^{1/4}\sim a m_s$, we see that the criterion for being able to
safely expand the DBI action into standard form is $\dot\phi^2 \ll
T_3 a^4 = V_{\rm inf}/( 8\pi^3 g_s)$.   On the other hand, the
slow-roll equation of motion gives $3 H\dot\phi \simeq -V_\phi$.  Since
$3H^2 \simeq V_{\rm inf}$, this
condition can thus be rewritten in terms of the slow-roll parameter
$\epsilon = \frac12(V_\phi/V)^2$, as
\be
	\epsilon \ll {3\over {16 \pi^3 g_s}}.
\ee
This is clearly satisfied in the present model, since as we will show,
$\epsilon\simeq 10^{-9}$ at the horizon crossing.

\subsection{Initial conditions}
\label{overshoot}

A potentially problematic aspect of the model is the need for special
initial conditions, even if the potential itself is not finely tuned. 
Obviously, inflation takes place near the inflection point of the
potential, so $\phi$ must not start lower than this point.  But  as
was pointed out in Ref.\ \cite{Underwood:2008dh}, $\phi$ also should
not start too much above the inflection point, because of the
overshoot problem: the inflaton can gain so much speed that it
quickly rolls past the inflection point without ever rolling slowly.
For the optimal parameter set we consider, however, there is another
consideration which prevents us from exploring the regime where
overshoot would take place.  This is because of the angular
directions of the extra dimensions in the throat, which we have set
to the values which minimize their potential.  As shown in Ref.\
\cite{Baumann}, the positions of the angular minima flip when $\phi$
exceeds a certain critical value $\phi_{\rm c}$, given by Eq.\ (C.24) of
that paper.  We would need to follow all the angular fields as well
to investigate this regime quantitatively, which is beyond the scope
of the present work. If the potential is assumed to have the same
form for  $\phi>\phi_{\rm c}$, we do observe overshooting, starting from
initial conditions of order $10\phi_{\rm c}$, but since we do not trust the
potential in this regime, no reliable statement about overshooting 
can be made in the present context.  

\begin{figure}[htp]
\centering{\includegraphics[width=4in]{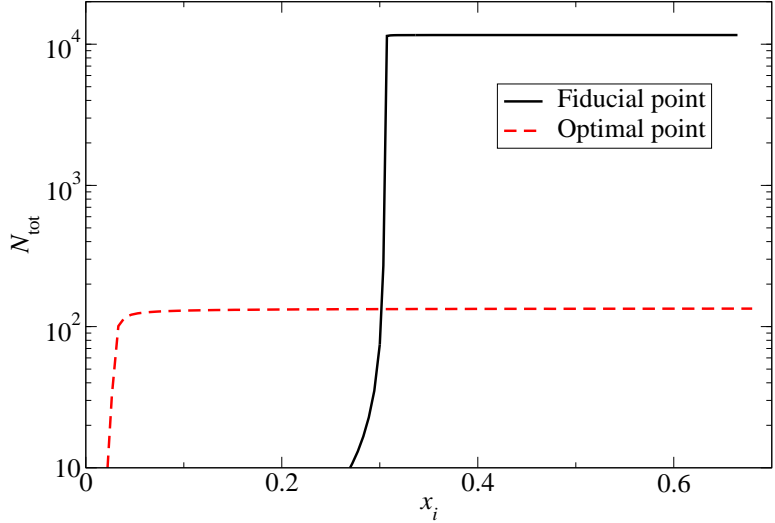}}
\caption{The total number of $e$-foldings as a function of the initial
condition $x_i$. The solid curve is for the fiducial parameters
(\ref{fid}), the dashed curve is for the optimal ones (\ref{optimal}).}
\label{Ntot_xi}
\end{figure}

Nevertheless, we can quantify the range of initial conditions over
which we get sufficient inflation and the potential is also valid:
the allowed initial separation of the branes is $x=0.301 \to 0.665$
for the fiducial point, while it is $x = 0.029 \to 0.680$ for the
optimal point.  We see that the allowed field range is expanded by a
factor of 2 in the optimal case. \Fig{Ntot_xi} shows the total number
of $e$-foldings as a function of the initial conditions in these two
cases. In an upcoming paper \cite{UC}, we will give a more
satisfactory solution to the problem of initial conditions in this
model, based on the idea of Ref.\ \cite{CS}.  

\subsection{One field versus two: the spectral index}

\begin{figure}[htp]
\centerline{\includegraphics[width=0.5\hsize]{\figdir 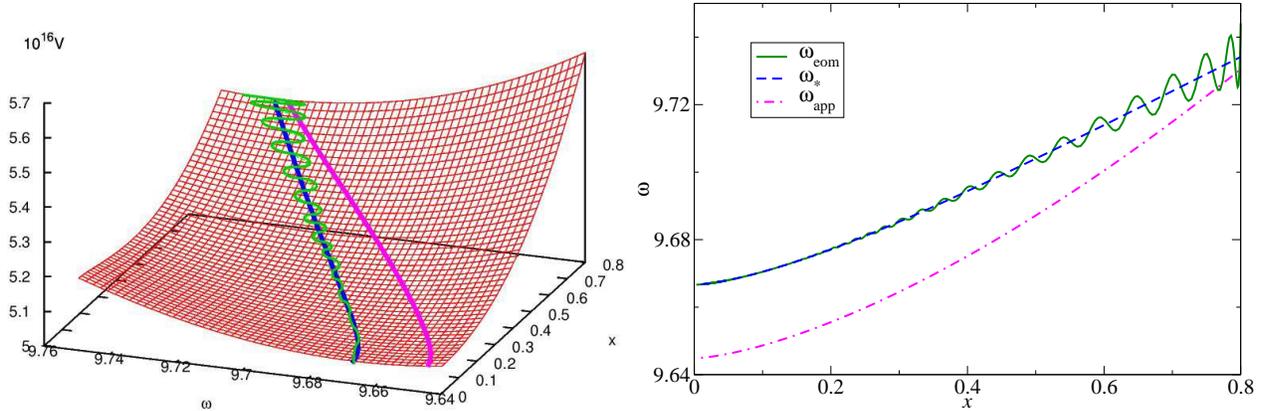}
\includegraphics[width=0.5\hsize]{\figdir traj9.eps}}
\caption{Left panel: actual inflationary trajectories in the valley-shaped potential,
and a plot of approximate trajectory \eq{wapp}.  
Right panel: projection of the same trajectories onto the $\omega$-$x$ plane.}
\label{traj}
\end{figure}

The present model has only one flat direction, so it is effectively a
single-field inflation model.  However, there is significant bending in the field
space of $\omega$ and $\phi$, as illustrated in  \fig{traj}. In the
left panel of \fig{traj}, 
the solid, smooth, leftmost curve is the ``instantaneous minimum"
$\omega_\ast(x)$, defined in \eq{wstar},  while the wavy 
curve is the actual trajectory found 
by solving the equations of motion, given some initial displacement
of the heavy field $\omega$ away from its instantaneous minimum.  We
used the initial condition $\omega_i = \omega_\ast(x_i)+0.01$, so
there are oscillations at first which allow one to distinguish
the two curves. The solid curve on the right is the effective
single-field description given by Refs.\ \cite{Baumann:2007np,
Baumann}:
\be
  \label{wapp}
  \omega \simeq \omega_0 \left[ 1+{1\over n\omega_F}\left(1-{1\over2\omega_F}\right)x^{3/2}\right].
\ee
As can be seen from the figure, it is not a very good approximation. Panda
{\it et al}.\ \cite{Panda} argued that using \eq{wapp}
underestimates the total number of $e$-foldings by an order of
magnitude. However, this does not invalidate the single-field
description; the real instantaneous minimum $\omega_\ast(x)$
does give a good approximation to the actual trajectory, once the
oscillations have Hubble damped away.  

However a subtlety can arise in the computation of the spectral
index when we use the slow-roll approximation
(\ref{nssr}).  The slow-roll parameters along the adiabatic direction
$\psi$ are related to those along the component field directions
$\phi\equiv \phi_\mu x$ and  $\chi\equiv\sqrt{3/2}\ln\omega$ by \eqs{epssr}{etasr} which we repeat here for convenience:\footnote{As
explained in Appendix \ref{eqs}, the slow-roll approximation is valid for
both component fields since the inflaton itself is rolling slowly.}  
\be
\eps_\psi \simeq \eps_\phi + \eps_\chi,\quad \eta_{\psi\psi}\simeq {1 \over \eps_\phi + \eps_\chi} \left(\eps_\phi\eta_{\phi\phi}
+  \eps_\chi\eta_{\chi\chi}+ {V_\phi V_\chi \over V^2}
\eta_{\phi\chi}\right).
\label{sreqs}
\ee
The problem occurs if we
try to use these expressions to compute the spectral index $n_s$ by
assuming the inflaton rolls exactly along the instantaneous minimum
$\omega = \omega_*(x)$.  By definition,
$\epsilon_\chi = \epsilon_\omega = 0$ along this trajectory. 
However, if we neglect the terms proportional to $\eps_\chi$ in 
$\eta_{\psi\psi}$, we get a result which does not agree with
computing the spectral index directly from $d\ln P/d\ln k$, \eq{ns}.

\begin{figure}[htp]
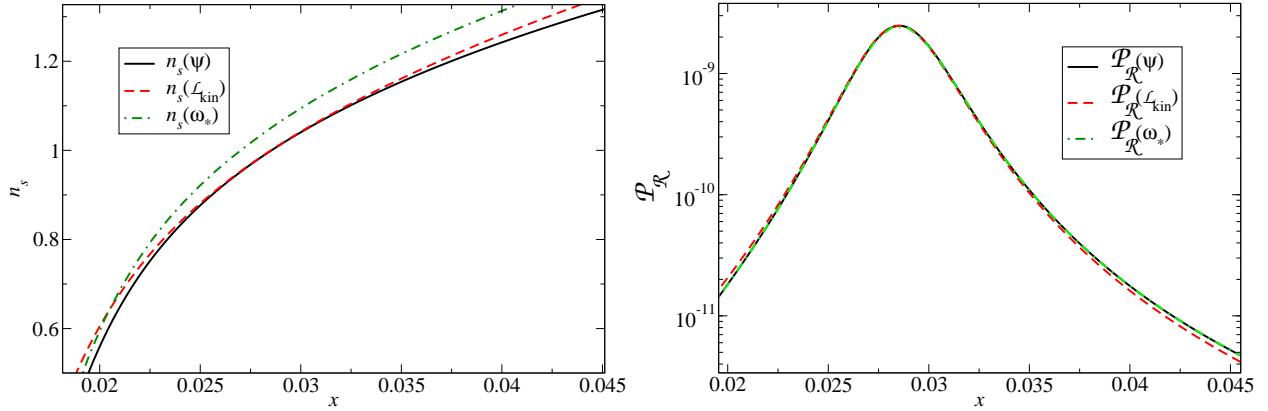

\centerline{\includegraphics[width=0.5\hsize]{\figdir ns5_2.eps}
\includegraphics[width=0.5\hsize]{\figdir Pk4_2.eps}}
\caption{Left panel: three approximations (described in the text) for the spectral index
as a function of inflaton position $x$, 
indicating a problem with the slow-roll approximation on the instantaneous minimum trajectory,
labeled $n_s(\omega_\ast)$.  Right panel: three approximations for 
the power spectrum as a function of $x$ (see text).}
\label{Pk}
\end{figure}

These conflicting results can be seen as the topmost (dot-dashed)  and
middle (dashed) curves, respectively, of \fig{Pk} (left panel),
labeled as $n_s(\omega_*)$ and $n_s({\cal L}_{\rm kin})$.  The latter is
based upon the approximation (\ref{prkin}) for the power spectrum,
and since it comes from directly differentiating  ${\cal P_R}(k)$, it must be
the correct result.  The curve $n_s(\omega_*)$ significantly
overestimates the spectral index as $n_s = 1.065$  near the
inflection point ($x=0.029$), whereas the correct value is $n_s =
0.989$ for this example.  Three approximations for the power spectrum
itself, as a function of inflaton position $x$, are shown in the
right panel of \fig{Pk}: ${\cal P}_{\cal R}(\psi)$, based on
the slow-roll approximation (\ref{prsr}), ${\cal P}_{\cal R}({\cal L}_{\rm
kin})$ using (\ref{prkin}) along the exact trajectory, and ${\cal
P}_{\cal R}(\omega_*)$ using  (\ref{prsr}) along the instantaneous
minimum $\omega = \omega_*(x)$. It can be seen that they all agree
quite well with each other near the inflection point, showing that
$\omega = \omega_*(x)$ is indeed a good approximation.  

\begin{figure}[htp]
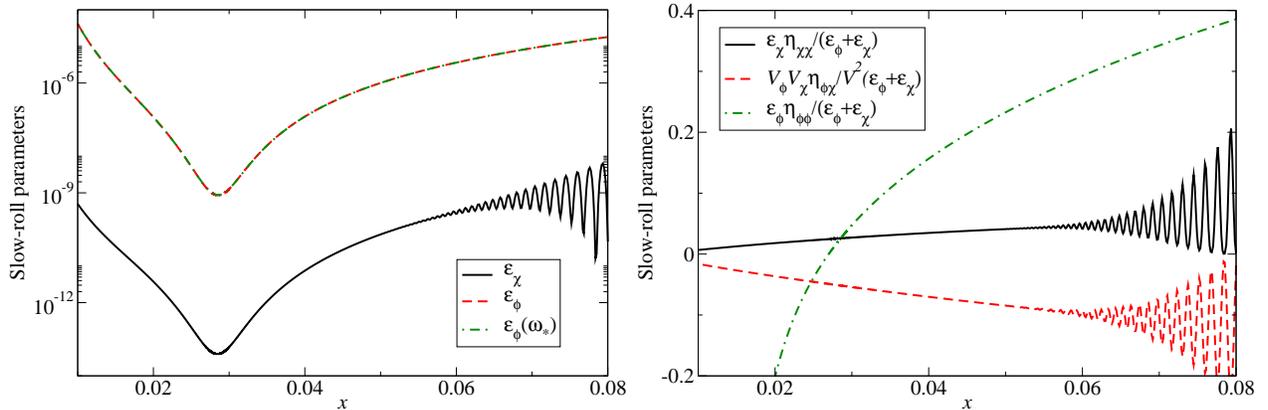

\centerline{\includegraphics[width=0.5\hsize]{\figdir eps4_2.eps}
\includegraphics[width=0.5\hsize]{\figdir eta4_3.eps}}
\caption{Left panel: the slow-roll parameters $\eps_\phi$ and $\epsilon_
\chi$.  $\epsilon_\phi$ is shown for both the exact trajectory and
the approximation $\omega = \omega_*(x)$.  Notice that $\eps_\chi=
0$ in that approximation, so $\eps_\chi$ is only shown for the exact
trajectory.  Right panel: the three terms which contribute to
$\eta_{\psi\psi}$ in \eq{sreqs}.}
\label{eps}
\end{figure}

Nevertheless, the resolution of the problem is that the actual inflaton trajectory
does not exactly follow the instantaneous minimum $\omega =
\omega_*(x)$; it deviates slightly from this, like a race car on a
banked curve.  Therefore $\epsilon_\chi\neq 0$ on the true trajectory
(although it is much smaller than $\epsilon_\phi$), as plotted in the
left panel of \fig{eps},  and the two terms
$\eps_\chi\eta_{\chi\chi}$ and $V_\phi V_\chi\eta_{\phi\chi}/V^2$ make an important contribution to 
$\eta_{\psi\psi}$.  In fact, the right panel of \fig{eps} shows
that these two terms very nearly cancel $\eps_\phi\eta_{\phi\phi}$ ($1-n_s=0.011$).  The
spectral index evaluated along the actual trajectory, but using the
slow-roll formula  (\ref{nssr}), is denoted $n_s(\psi)$. \Fig{Pk} shows that it gives a good fit to the numerically computed
index $n_s({\cal L}_{\rm kin})$ in the most important region, where most of
inflation is taking place.  Therefore the two-field slow-roll formula
for $n_s$ is a good approximation, but only if one uses the correct
two-field trajectory and not the instantaneous minimum approximation.

\section{Conclusions}

In this paper we have made a detailed study of the warped
$D3$-$\overline{D3}$ inflation model, accounting for superpotential
corrections, from a phenomenological perspective, but also with
attention to the need for  self-consistency from the string
theoretical point of view.   We extended the model slightly by
including uplifting from throats other than the inflationary one,
which was necessary so that inflation could end with a nearly
vanishing cosmological constant.   We subsequently explored the
parameter space using Monte Carlo methods, finding for the first
time values which satisfy all theoretical and experimental
constraints.   Moreover we identified an optimal set of parameters in
the vicinity of which there appears to be no need for fine-tuning.
This arises in part  because of a peculiar feature of the potential in this
model: there are ranges of the parameter $D_{01}$ for which one finds a {\it
minimum} number of $e$-foldings of inflation, because at the boundaries
of such regions, a local minimum develops, giving rise to
$N_{\rm tot}\to\infty$.  Toward this end, we defined the concept of a
relative volume $\delta_N$ in an $N$-dimensional parameter space
consistent with successful inflation, and having the property that  
$\sqrt[N]{\delta_N}$ represents the average degree of fine-tuning of any
parameter.  We obtained an improvement by 8 orders of magnitude in 
$\delta_4$ for the 4-parameter subspace which was needed to fully
describe the range of potential shapes in the model, relative to the
fine-tuned example from the literature with which we started our
Monte Carlo search.   We conclude that this string theoretic inflation
model is not as delicate as it first seemed in Ref.\
\cite{Baumann:2007np}.

There are some caveats to this successful conclusion however.  The
value of the K\"ahler modulus at the optimal point, $\omega_F=9.6$,
is only marginally large enough to give one confidence that the
low-energy effective field theory is not significantly perturbed by
higher dimensional operators (coming from integrating out the extra
dimensions) which are supposed to be suppressed by large $\omega_F$.
On the other hand, since the scale of the potential goes like
$e^{-2\omega_F} A_0^2/M_{\rm Pl}^2$, the COBE normalization demands that 
increasing $\omega_F$ must be accompanied by an exponential increase
in $A_0$.  The latter is related to the energy scale $\Lambda$ of gaugino
condensation (or a Euclidean D3-brane) by $A\sim\Lambda^3$, which
should certainly not exceed the Planck scale, and likely should also
lie below the warped string scale.  This could be a source of
theoretical tension for the model.  

On the phenomenological side, since we allowed for 2$\sigma$
deviations of $n_s$ from the WMAP5 central value in our Monte Carlo
search, we can detect a statistical preference of the model for
larger values of $n_s$ near $0.99$.  That is to say, the least
fine-tuned models which we find correspond to such higher values of
$n_s$, and therefore if future data are shown to prefer lower values,
it would be an indication disfavoring the model.  In this way, our
global search of the parameter space helps to provide an additional
predictive tool which would not be available by simply finding a few
sets of parameter values which were consistent with the data.  This
statement assumes that there is a landscape of string vacua which
allows for nature to scan through the possible values of the
parameters.

We also investigated an alternative measure of the
degree of fine-tuning, i.e., the absolute allowed volume.
While the motivation for this measure is different from that of the
relative measure, the result is nearly the same---we found regions in
the parameter space where the absolute allowed volume is increased by
8 orders of magnitude, and the overall degree of tuning per
parameter is still at the 10\% level. This finding reinforces our
point that parameter regions exist where the severity of
fine-tuning is greatly ameliorated.

{\bf Note added}.  While we were finishing this work, Ref.\
\cite{Kachru} appeared, where it was pointed out that generic
deformations of the throat geometry can have a qualitatively similar
effect to the superpotential corrections, in allowing for an
inflection point in the potential.  More recently  Ref.\  \cite{Ali}
appeared, which reaches a different conclusion than ours, finding
too large a value of $n_s$.  Their work starts with a somewhat
different model, based on the deformations discussed in Ref.\ \cite{Kachru}
rather than superpotential corrections.  But since the latter model is
supposed to have qualitatively similar behavior to the former, 
the result of Ref.\ \cite{Ali} looks surprising.  This discrepancy
comes from the neglect of the Coulomb interaction in Ref.\ \cite{Ali}.
We thank the authors of Ref.\ \cite{Baumann} for clarifying this
point for us.

\section*{Acknowledgments} J.~C.\ thanks the Banff International Research
Station, where this work was started.  
 We thank Bret Underwood for helpful
discussions.  L.~H.\ is supported by
Carl Reinhardt Fellowship at McGill University.
Our research is also supported by NSERC (Canada).

\appendix

\section{Equations}
\label{eqs}

Here we compile various  formulas needed in the previous sections.

\subsection{$\omega_\ast(x)$ and $\omega_0$}
\label{omega*}
The instantaneous minimum $\omega_\ast (x)$ is defined through \eq{wstar}.  This leads to the result
\bea
  \label{w*}
  &&2s\omega_Fe^{2\omega_\ast-2\omega_F}g^{-2/n}\left(1+\frac{D_{01}}{1+C_D\begin{displaystyle}{D_0 \over x^4}\end{displaystyle}}\right)-(2\omega_F+3)\left(2+\omega_\ast-{1\over6}\omega_0\phi_\mu^2x^2\right)e^{\omega_\ast-\omega_F}g^{-1/n}\nonumber\\
  &&+2\omega_\ast^2+\left[7-{1\over3}\omega_0\phi_\mu^2x^2+{3 \over ng}\left({cx \over g}-x^{3/2}\right)\right]\omega_\ast +{3 \over ng}\left({cx \over g}-x^{3/2}\right)\left(1-{1\over6}\omega_0\phi_\mu^2x^2\right)\nonumber\\
  &&-{5\over6}\omega_0\phi_\mu^2x^2+6=0.
\eea
Setting $x=0$, we have the equation for $\omega_0 = \omega_0(s,\omega_F)$:
\be
  \label{w0}
  2s\omega_Fe^{2\omega_0-2\omega_F}-(2\omega_F+3)(2+\omega_0)e^{\omega_0-\omega_F}+2 \omega_0^2 +7\omega_0+6=0.
\ee
This equation generically has two solutions for $\omega_0$; 
the one closest to $\omega_F$ is the minimum while the one farther
from $\omega_F$ is at the maximum of the potential.  

If we apply the uplifting condition, $V(0,\omega_0)=0$, then the parameter $s$ is fixed and $\omega_0$ is a function of $\omega_F$ only; see Section \ref{uplifting}.

\subsection{Microscopic constraints on the parameters}
\label{microscopic}
There is a constraint on the allowed field range of the inflaton field \cite{Baumann}:
\be
  \Delta\phi < {2 \over \sqrt{N_5}}.
\ee
If we consider $0<x<1$ only, then the parametrization \eq{phimu} requires
$B_6 > 1$ and $Q_\mu > 1$.  It is also required that 
$B_4 > 1.$
To satisfy the COBE normalization, one requires 
$\omega_0 < O(30)$; otherwise the inflation scale will be too low.
Using the fact that  $\omega_F\simeq\omega_0$ leads to 
the further constraint ${N_5/n} < O(10^2)$.

\subsection{Potential and its derivatives}
\label{potapp}
We rewrite the potential as
\be
  V = V_F + V_D \equiv {a|A_0|^2 \over 3U^2} (V_f + V_d),
\ee
where
\bea
  V_f &=& e^{-2\omega}g^{2/n}\left[2\omega+6-2(2\omega_F+3)e^{\omega-\omega_F}g^{-1/n} + {3\over ng}\left({cx\over g}-x^{3/2}\right)\right],\\
  V_d &=& 2s\omega_Fe^{-2\omega_F}\left(1+\frac{D_{01}}{1+C_D\begin{displaystyle}{D_0 \over x^4}\end{displaystyle}}\right).
\eea
Then we have
\bea
  {\partial V\over\partial\omega} &=& {a|A_0|^2 \over 3U^2} \left\{-{4(V_f+V_d) \over aU} - 2V_f + 2\left[1-(2\omega_F+3)e^{\omega-\omega_F}g^{-1/n}\right]e^{-2\omega}g^{2/n}\right\},\\
  {\partial V\over\partial x} &=& {a|A_0|^2 \over 3U^2} \left[{4\omega_0\phi_\mu^2x\over3aU}(V_f+V_d)+{\partial V_f \over \partial x} + {\partial V_d \over\partial x}\right],
\eea
where
\bea
  {\partial V_f \over \partial x} &=& {3\over ng}\left\{\sqrt{x}V_f + e^{-2\omega}g^{2/n}\left[\sqrt{x}(2\omega_F+3)e^{\omega-\omega_F}g^{-1/n}-{3\sqrt{x}\over g}\left(\hf+{cx\over g}\right)+{c\over g}\right]\right\},\\
  {\partial V_d \over \partial x} &=& 8s\omega_Fe^{-2\omega_F}C_DD_1D_{01}^2x^{-5}\left(1+{C_DD_0\over x^4}\right)^{-2}.
\eea

We need the second order derivatives when calculating the slow-roll 
parameters:
\bea
  {\partial^2 V \over \partial\omega^2} &=& -{8 \over aU} \left( {V \over aU} + {\partial V \over \partial\omega}\right) + {2a|A_0|^2 \over 3U^2} \left\{2V_f + e^{-2\omega}g^{2/n}\left[-4+3(2\omega_F+3)e^{\omega-\omega_F}g^{-1/n}\right]\right\},\nonumber\\
  &&\\
  {\partial^2 V \over \partial x^2} &=& {4\omega_0\phi_\mu^2 \over 3aU}\left(V+2x{\partial V\over\partial x}-{2\omega_0\phi_\mu^2x^2\over 3aU}V\right) + {a|A_0|^2 \over 3U^2}\left({\partial^2V_f \over \partial x^2} + {\partial^2V_d \over \partial x^2}\right),
\eea
where
\bea
  {\partial^2V_f \over \partial x^2} &=& {3\sqrt{x}\over ng} \left\{\left({1\over2x}-{3\sqrt{x}\over ng}\right)V_f+\left(2-{n\over2}\right){\partial V_f \over \partial x} + e^{-2\omega}g^{2/n}\left[(2\omega_F+3)e^{\omega-\omega_F}g^{-1/n}\right.\right.\nonumber\\
  &&\left.\left.\times\left({1\over2x} - {3\sqrt{x}\over2ng}\right)+{3\over2g}\left({6cx^{3/2}\over g^2}+{3\sqrt{x}\over2g}-{3c\over g}-{1\over2x}\right)-{3c\over2g^2}\right]\right\},\\
  {\partial^2V_d \over\partial x^2} &=& -40s\omega_F{\rm
e}^{-2\omega_F}C_DD_1D_{01}^2x^{-6}\left(1+{C_DD_0\over
x^4}\right)^{\!\!-2}\!\left[1-{8C_DD_0\over5x^4}\left(1+{C_DD_0\over x^4}\right)^{-1}\right],
\eea
and
\bea
  {\partial^2V\over\partial x\partial\omega} &=& {4\omega_0\phi_\mu^2x \over 3aU}\left[2\left(1+{1\over aU}\right)V+{\partial V\over\partial\omega}\right]-2\left(1+{2\over aU}\right){\partial V\over\partial x}\nonumber\\
  &&+ {2a|A_0|^2\over3U^2}\left\{{\partial V_d\over\partial x}+{3\sqrt{x}\over2ng}e^{-2\omega}g^{2/n}\left[2-(2\omega_F+3)e^{\omega-\omega_F}g^{-1/n}\right]\right\}.
\eea

\subsection{Power spectrum}
\label{pwspt}
The kinetic part of the Lagrangian is
\be
  {\cal L}_{\rm kin} = {a^2U^2 \over 4}\left({1 \over 3}\pi_\omega^2 + \frac{\pi_x^2}{2\omega_0\omega\phi_\mu^2}\right).
\ee
Its derivative is given by
\be
  \frac{d{\cal L}_{\rm kin}}{dN} = {4{\cal L}_{\rm kin} \over aU}\left({d\omega \over dN} - {\omega_0\phi_\mu^2x \over 3}{dx\over dN}\right) + {a^2U^2\over 4}\left[{2\pi_\omega \over 3}{d\pi_\omega \over dN} + \frac{\pi_x}{\omega_0\omega\phi_\mu^2}\left({d\pi_x\over dN}-{\pi_x \over 2\omega}{d\omega \over dN}\right)\right],
\ee
where $dx/dN$, $d\omega/dN$, $d\pi_x/dN$, and $d\pi_\omega/dN$ are obtained by solving the equations of motion. From \eq{fieldmann}, we have
\be
  {d\ln H \over dN} = {1 \over 6H^2} \left({d{\cal L}_{\rm kin} \over dN} + V_x{dx \over dN} + V_\omega{d\omega \over dN}\right).
\ee
The spectral index can be calculated through \eq{ns}.

\subsection{Slow-roll parameters}
\label{srapp}
The slow-roll parameters for a generic field $\phi$ are defined as
\bea
  \eps_\phi &=& {1 \over 2}\left({V_\phi \over V}\right)^2,\\
  \eta_{\phi\phi} &=& {V_{\phi\phi} \over V},
\eea
where $V_{\phi\phi} \equiv {\partial^2 V /\partial\phi^2}$.
Since \cite{Gordon:2000hv}
\bea
  V_\psi &=& V_\phi\cos\theta + V_\chi\sin\theta,\\
  V_{\psi\psi} &=& V_{\phi\phi}\cos^2\theta + V_{\chi\chi}\sin^2\theta + V_{\phi\chi}\sin2\theta,
\eea
we have
\bea
  \eps_\psi &=& {1 \over \dot\phi^2 + \dot\chi^2}\left(\dot\phi^2\eps_\phi + \dot\chi^2\eps_\chi + \dot\phi\dot\chi{V_\phi V_\chi\over V^2}\right),\\
  \label{etapp}
  \eta_{\psi\psi} &=& {1 \over \dot\phi^2 + \dot\chi^2}\left(\dot\phi^2\eta_{\phi\phi} + \dot\chi^2\eta_{\chi\chi} + 2\dot\phi\dot\chi\eta_{\phi\chi}\right).
\eea
In general, one should not apply the slow-roll approximation to all the
terms in these expressions since one linear combination of the fields
is heavy.  However, we are interested in trajectories along the flat
direction of the potential, after any oscillations in the steep
directions have Hubble damped away.  In this case, the slow-roll
approximation is valid along both of the field components, and we can write
\bea
  \label{epssr}
  \eps_\psi &\simeq& \eps_\phi + \eps_\chi,\\
  \label{etasr}
  \eta_{\psi\psi} &\simeq& {1 \over \eps_\phi + \eps_\chi}\left(\eps_\phi\eta_{\phi\phi} + \eps_\chi\eta_{\chi\chi}+ {V_\phi V_\chi \over V^2}\eta_{\phi\chi}\right).
\eea

Using the relation between $(x,\omega)$ and $(\phi,\chi)$, we have
\bea
  \eps_\phi &=& {1\over\phi_\mu^2}\eps_x,\\
  \eps_\chi &=& {2\over3}\omega^2\eps_\omega,\\
  \eta_{\phi\phi} &=& {1\over\phi_\mu^2}\eta_{xx},\\
  \eta_{\chi\chi} &=& {2\over3}\omega\left({V_\omega\over V}+\omega\eta_{\omega\omega}\right),\\
  \eta_{\phi\chi} &=& \sqrt{2\over3}{\omega\over\phi_\mu}\eta_{x\omega}.
\eea
The fact that the slow-roll approximation for the spectral index
agrees with the numerical computation of $n_s$ provides further
evidence for the validity of the approximation.  

\section{Objective Function for Monte Carlo Method}
\label{empirical}

In Section \ref{metro} we described the Metropolis algorithm  for
finding parameters which maximize the relative volume.  The strategy
is to minimize an objective function. A naive choice of the
objection could be the negative relative volume:
\be
  f_{\rm obj} = - \delta_N.
\ee
However, the concept of the relative volume is only valid when the
potential is monotonic, so the objective function becomes zero
in the nonmonotonic regions,
\be
  f_{\rm obj} = \left\{
  \begin{array}{ll}
    0,&\hskip0.5in V\ \hbox{nonmonotonic},\\
    -\delta_N,&\hskip0.5in\delta_N>0.
  \end{array}\right.
\ee

But this is not a useful choice, since most points in the
$N$-dimensional space give nonmonotonic potentials. Supposing that we
start with an $N$-dimensional simplex, if we do not choose the $N+1$
initial vertices carefully, then there is a good chance that all
$N+1$ vertices correspond to nonmonotonic potentials. Therefore all
the initial vertices have the same value of the objective function,
and it may take a long time for the code to escape from such a
region, since, apparently, there is no downhill direction. This
problem might be solved by selecting good initial vertices;
however, it is not practical because, before first getting a few
successful chains, one does not know the configuration of the
$N$-dimensional parameter space.

To avoid this situation, we need another objective function for the
nonmonotonic regime. The primary characteristic of a nonmonotonic
potential, having a local minimum, is that the number of
$e$-foldings diverges as the inflaton gets stuck in the minimum. 
However, in the Runge-Kutta method (with adaptive stepsize control)
which we use to solve the inflaton equations of motion, a parameter
{\tt hmin} controls the minimum stepsize, which causes the evolution
to end after a finite number of $e$-foldings.  This number will tend
to be larger for a shallow local minimum than for a deep one.  Thus,
to assist the program in finding parameters that move away from a 
local minimum, a good choice for the objective function in the
nonmonotonic regime is  
\be
  f_{\rm obj} = \left\{
  \begin{array}{ll}
    N^\prime-N_{\rm tot},&\hskip0.5in V\ \hbox{nonmonotonic},\\
    -\delta_N,&\hskip0.5in\delta_N>0.
  \end{array}\right.
\ee
where $N^\prime$ is some
large number which is not supposed to be attained in the code.

The above objective function should work. However, we not only want
the central point to give a monotonic potential, but we also need it
to satisfy the experimental cosmic microwave background constraints. The next step is to find
the value of the inflaton field $x$ where the COBE normalization is
satisfied ${\cal P_R}(x_{\rm COBE}) = 2.41\times10^{-9}$. (Except for
the case whose whole power spectrum is less than the COBE
normalization scale, this point $x_{\rm COBE}$ can always be found.)
To make this correspond to the right scale of wave numbers, we also
need the number of $e$-foldings at this field value to satisfy
$50\le N_{\rm COBE}\le60$. This leads to the further refinement:
\be
  f_{\rm obj} = \left\{
  \begin{array}{ll}
    N^\prime-N_{\rm tot},&\hskip0.5in V\ \hbox{nonmonotonic},\\
    50-N_{\rm COBE},&\hskip0.5in N_{\rm COBE}<50,\\
    N_{\rm COBE}-60,&\hskip0.5in N_{\rm COBE}>60,\\
    -\delta_N,&\hskip0.5in\delta_N>0.
  \end{array}\right.
\ee

If the constraint on $N_{\rm COBE}$ is satisfied, then the next step
is the constraint on the spectral index. We require  it to
be within 2$\sigma$ of WMAP5's mean value, so we add
\be
  f_{\rm obj} = \left\{
  \begin{array}{ll}
    N^\prime-N_{\rm tot},&\hskip0.5in V\ \hbox{nonmonotonic},\\
    50-N_{\rm COBE},&\hskip0.5in N_{\rm COBE}<50,\\
    N_{\rm COBE}-60,&\hskip0.5in N_{\rm COBE}>60,\\
    |n_s-0.963|,&\hskip0.5in 50\le N_{\rm COBE}\le60,\\
    -\delta_N,&\hskip0.5in 0.935\le n_s\le0.991.
  \end{array}\right.
\ee

Finally,  we must avoid overlapping conditions 
 between the above cases.   We thus modify it to
\be
  f_{\rm obj} = \left\{
  \begin{array}{ll}
    {\tt max}(50,N^\prime-N_{\rm tot}),&\hskip0.5in V\ \hbox{nonmonotonic},\\
    {\tt max}(2,50-N_{\rm COBE}),&\hskip0.5in N_{\rm COBE}<50,\\
    {\tt min}(50,{\tt max}(2,N_{\rm COBE}-60)),&\hskip0.5in N_{\rm COBE}>60,\\
    |n_s-0.963|,&\hskip0.5in 50\le N_{\rm COBE}\le60,\\
    -\delta_N,&\hskip0.5in 0.935\le n_s\le0.991,
  \end{array}\right.
\ee
where the number 50 is the minimum number of e-folding we need, while
the number 2 is somewhat arbitrary (we suppose that 
$|n_s-0.963|<2$).

\end{document}